\documentclass[journal]{IEEEtran}
\IEEEoverridecommandlockouts
\usepackage[most]{tcolorbox}
\usepackage{amsmath}
\usepackage{amsthm}
\usepackage{breakurl}
\usepackage{url}
\usepackage{tikz}
\usepackage{multirow}
\usepackage{algpseudocode}
\usepackage{algorithm}

\usepackage{colortbl}
\usepackage[nolist]{acronym}

\usepackage{graphicx}
\usepackage{xcolor}
\usepackage{color}

\usepackage{varwidth}
\usepackage{rotating}

\usepackage{nameref}
\usepackage{zref-xr,zref-user}
\zxrsetup{toltxlabel=true, tozreflabel=false}
\usepackage{cite}

\usepackage{tikz-3dplot}

\usepackage{subfigure}

\usetikzlibrary{decorations.pathreplacing,decorations.markings}

\tikzset{arrow data/.style 2 args={%
      decoration={%
         markings,
         mark=at position #1 with \arrow{#2}},
         postaction=decorate}
      }%

\usetikzlibrary{shapes,backgrounds,calc}

\tikzstyle{dummy} = [rectangle, text width=0.1em, draw=white, white,
                      minimum width=0.1em, minimum height=3em, opacity=0.0]

\makeatletter
\newcommand*{\Strut}[1][0.1em]{\vrule\@width\z@\@height#1\@depth\z@\relax}
\makeatother

\definecolor{Linen}{rgb}{0.9803,0.9411,0.9019}
\definecolor{White}{rgb}{1,1,1}
\definecolor{Coral}{rgb}{1,0.4980,0.3137}
\definecolor{Grayblue}{rgb}{0.9411,0.9411,0.9803}
\definecolor{DarkLinen}{rgb}{0.729,0.7176,0.635}

\begin{document}

\begin{acronym}

\acro{5G-NR}{5G New Radio}
\acro{3GPP}{3rd Generation Partnership Project}
\acro{AC}{address coding}
\acro{ACF}{autocorrelation function}
\acro{ACR}{autocorrelation receiver}
\acro{ADC}{analog-to-digital converter}
\acrodef{aic}[AIC]{Analog-to-Information Converter}     
\acro{AIC}[AIC]{Akaike information criterion}
\acro{aric}[ARIC]{asymmetric restricted isometry constant}
\acro{arip}[ARIP]{asymmetric restricted isometry property}

\acro{ARQ}{automatic repeat request}
\acro{AUB}{asymptotic union bound}
\acrodef{awgn}[AWGN]{Additive White Gaussian Noise}     
\acro{AWGN}{additive white Gaussian noise}
\acro{AoA}{Angle of Arrival}

\acro{APSK}[PSK]{asymmetric PSK} 

\acro{waric}[AWRICs]{asymmetric weak restricted isometry constants}
\acro{warip}[AWRIP]{asymmetric weak restricted isometry property}
\acro{BCH}{Bose, Chaudhuri, and Hocquenghem}        
\acro{BCHC}[BCHSC]{BCH based source coding}
\acro{BEP}{bit error probability}
\acro{BFC}{block fading channel}
\acro{BG}[BG]{Bernoulli-Gaussian}
\acro{BGG}{Bernoulli-Generalized Gaussian}
\acro{BPAM}{binary pulse amplitude modulation}
\acro{BPDN}{Basis Pursuit Denoising}
\acro{BPPM}{binary pulse position modulation}
\acro{BPSK}{binary phase shift keying}
\acro{BPZF}{bandpass zonal filter}
\acro{BSC}{binary symmetric channels}              
\acro{BU}[BU]{Bernoulli-uniform}
\acro{BER}{bit error rate}
\acro{BS}{Base Station}

\acro{CP}{Cyclic Prefix}
\acrodef{cdf}[CDF]{cumulative distribution function}   
\acro{CDF}{cumulative distribution function}
\acrodef{c.d.f.}[CDF]{cumulative distribution function}
\acro{CCDF}{complementary cumulative distribution function}
\acrodef{ccdf}[CCDF]{complementary CDF}               
\acrodef{c.c.d.f.}[CCDF]{complementary cumulative distribution function}
\acro{CD}{cooperative diversity}

\acro{CDMA}{Code Division Multiple Access}
\acro{ch.f.}{characteristic function}
\acro{CIR}{channel impulse response}
\acro{cosamp}[CoSaMP]{compressive sampling matching pursuit}
\acro{CR}{cognitive radio}
\acro{cs}[CS]{compressed sensing}                   
\acrodef{cscapital}[CS]{Compressed sensing} 
\acrodef{CS}[CS]{compressed sensing}
\acro{CSI}{channel state information}
\acro{CCSDS}{consultative committee for space data systems}
\acro{CC}{convolutional coding}
\acro{Covid19}[COVID-19]{Coronavirus disease}
\acro{CAPEX}{CAPital EXpenditures}

\acro{DAA}{detect and avoid}
\acro{DAB}{digital audio broadcasting}
\acro{DCT}{discrete cosine transform}
\acro{dft}[DFT]{discrete Fourier transform}
\acro{DR}{distortion-rate}
\acro{DS}{direct sequence}
\acro{DS-SS}{direct-sequence spread-spectrum}
\acro{DTR}{differential transmitted-reference}
\acro{DVB-H}{digital video broadcasting\,--\,handheld}
\acro{DVB-T}{digital video broadcasting\,--\,terrestrial}
\acro{DL}{downlink}
\acro{DSSS}{Direct Sequence Spread Spectrum}
\acro{DFT-s-OFDM}{Discrete Fourier Transform-spread-Orthogonal Frequency Division Multiplexing}
\acro{DAS}{distributed antenna system}
\acro{DNA}{Deoxyribonucleic Acid}
\acro{DL-TDoA}{DownLink  Time Difference of Arrival}

\acro{EC}{European Commission}
\acro{EED}[EED]{exact eigenvalues distribution}
\acro{EIRP}{Equivalent Isotropically Radiated Power}
\acro{ELP}{equivalent low-pass}
\acro{eMBB}{Enhanced Mobile Broadband}
\acro{EMF}{Electro-Magnetic Field}
\acro{EU}{European union}
\acro{ELP}{Exposure Limit-based Power}
\acro{ECDF}{Empirical Cumulative Distribution Function}

\acro{FC}[FC]{fusion center}
\acro{FCC}{Federal Communications Commission}
\acro{FEC}{forward error correction}
\acro{FFT}{fast Fourier transform}
\acro{FH}{frequency-hopping}
\acro{FH-SS}{frequency-hopping spread-spectrum}
\acrodef{FS}{Frame synchronization}
\acro{FSsmall}[FS]{frame synchronization}  
\acro{FDMA}{Frequency Division Multiple Access}  
\acro{FSPL}{Free Space Path Loss}

\acro{GA}{Gaussian approximation}
\acro{GF}{Galois field }
\acro{GG}{Generalized-Gaussian}
\acro{GIC}[GIC]{generalized information criterion}
\acro{GLRT}{generalized likelihood ratio test}
\acro{GPS}{Global Positioning System}
\acro{GMSK}{Gaussian minimum shift keying}
\acro{GSMA}{Global System for Mobile communications Association}

\acro{HAP}{high altitude platform}

\acro{IDR}{information distortion-rate}
\acro{IFFT}{inverse fast Fourier transform}
\acro{iht}[IHT]{iterative hard thresholding}
\acro{i.i.d.}{independent, identically distributed}
\acro{IoT}{Internet of Things}                      
\acro{IR}{impulse radio}
\acro{lric}[LRIC]{lower restricted isometry constant}
\acro{lrict}[LRICt]{lower restricted isometry constant threshold}
\acro{ISI}{intersymbol interference}
\acro{ITU}{International Telecommunication Union}
\acro{ICNIRP}{International Commission on Non-Ionizing Radiation Protection}
\acro{IEEE}{Institute of Electrical and Electronics Engineers}
\acro{ICES}{IEEE international committee on electromagnetic safety}
\acro{IEC}{International Electrotechnical Commission}
\acro{IARC}{International Agency on Research on Cancer}
\acro{IS-95}{Interim Standard 95}

\acro{KPI}{Key Performance Indicator}

\acro{LEO}{low earth orbit}
\acro{LF}{likelihood function}
\acro{LLF}{log-likelihood function}
\acro{LLR}{log-likelihood ratio}
\acro{LLRT}{log-likelihood ratio test}
\acro{LOS}{Line-of-Sight}
\acro{LRT}{likelihood ratio test}
\acro{wlric}[LWRIC]{lower weak restricted isometry constant}
\acro{wlrict}[LWRICt]{LWRIC threshold}
\acro{LPWAN}{low power wide area network}
\acro{LoRaWAN}{Low power long Range Wide Area Network}
\acro{NLOS}{non-line-of-sight}
\acro{LB}{Lower Bound}
\acro{LCS}{LoCation Service}

\acro{MB}{multiband}
\acro{MC}{multicarrier}
\acro{MDS}{mixed distributed source}
\acro{MF}{matched filter}
\acro{m.g.f.}{moment generating function}
\acro{MI}{mutual information}
\acro{MIMO}{Multiple-Input Multiple-Output}
\acro{MISO}{multiple-input single-output}
\acrodef{maxs}[MJSO]{maximum joint support cardinality}                       
\acro{ML}[ML]{maximum likelihood}
\acro{MMSE}{minimum mean-square error}
\acro{MMV}{multiple measurement vectors}
\acrodef{MOS}{model order selection}
\acro{M-PSK}[${M}$-PSK]{$M$-ary phase shift keying}                       
\acro{M-APSK}[${M}$-PSK]{$M$-ary asymmetric PSK} 
\acro{MSP}{Minimum Sensitivity-based Power}

\acro{M-QAM}[$M$-QAM]{$M$-ary quadrature amplitude modulation}
\acro{MRC}{maximal ratio combiner}                  
\acro{maxs}[MSO]{maximum sparsity order}                                      
\acro{M2M}{machine to machine}                                                
\acro{MUI}{multi-user interference}
\acro{mMTC}{massive Machine Type Communications}      
\acro{mm-Wave}{millimeter-wave}
\acro{MP}{mobile phone}
\acro{MPE}{maximum permissible exposure}
\acro{MAC}{media access control}
\acro{NB}{narrowband}
\acro{NBI}{narrowband interference}
\acro{NLA}{nonlinear sparse approximation}
\acro{NLOS}{Non-Line of Sight}
\acro{NTIA}{National Telecommunications and Information Administration}
\acro{NTP}{National Toxicology Program}
\acro{NHS}{National Health Service}

\acro{OC}{optimum combining}                             
\acro{OC}{optimum combining}
\acro{ODE}{operational distortion-energy}
\acro{ODR}{operational distortion-rate}
\acro{OFDM}{orthogonal frequency-division multiplexing}
\acro{omp}[OMP]{orthogonal matching pursuit}
\acro{OSMP}[OSMP]{orthogonal subspace matching pursuit}
\acro{OQAM}{offset quadrature amplitude modulation}
\acro{OQPSK}{offset QPSK}
\acro{OFDMA}{Orthogonal Frequency-division Multiple Access}
\acro{OPEX}{OPerating EXpenditures}
\acro{OQPSK/PM}{OQPSK with phase modulation}

\acro{PAM}{pulse amplitude modulation}
\acro{PAR}{peak-to-average ratio}
\acrodef{pdf}[PDF]{probability density function}                      
\acro{PDF}{probability density function}
\acrodef{p.d.f.}[PDF]{probability distribution function}
\acro{PDP}{power dispersion profile}
\acro{PMF}{probability mass function}                             
\acrodef{p.m.f.}[PMF]{probability mass function}
\acro{PN}{pseudo-noise}
\acro{PPM}{pulse position modulation}
\acro{PRake}{Partial Rake}
\acro{PSD}{power spectral density}
\acro{PSEP}{pairwise synchronization error probability}
\acro{PSK}{phase shift keying}
\acro{PD}{Power Density}
\acro{8-PSK}[$8$-PSK]{$8$-phase shift keying}
\acro{PSL}{Positioning Service Level}

\acro{FSK}{frequency shift keying}

\acro{QAM}{Quadrature Amplitude Modulation}
\acro{QPSK}{quadrature phase shift keying}
\acro{OQPSK/PM}{OQPSK with phase modulator }

\acro{RD}[RD]{raw data}
\acro{RDL}{"random data limit"}
\acro{ric}[RIC]{restricted isometry constant}
\acro{rict}[RICt]{restricted isometry constant threshold}
\acro{rip}[RIP]{restricted isometry property}
\acro{ROC}{receiver operating characteristic}
\acro{rq}[RQ]{Raleigh quotient}
\acro{RS}[RS]{Reed-Solomon}
\acro{RSC}[RSSC]{RS based source coding}
\acro{RFP}{Radio Frequency ``Pollution''}
\acro{r.v.}{random variable}                               
\acro{R.V.}{random vector}
\acro{RMS}{root mean square}
\acro{RFR}{radiofrequency radiation}
\acro{RIS}{Reconfigurable Intelligent Surface}
\acro{RNA}{RiboNucleic Acid}

\acro{SA}[SA-Music]{subspace-augmented MUSIC with OSMP}
\acro{SCBSES}[SCBSES]{Source Compression Based Syndrome Encoding Scheme}
\acro{SCM}{sample covariance matrix}
\acro{SEP}{symbol error probability}
\acro{SG}[SG]{sparse-land Gaussian model}
\acro{SIMO}{single-input multiple-output}
\acro{SINR}{Signal-to-Interference plus Noise Ratio}
\acro{SIR}{signal-to-interference ratio}
\acro{SISO}{single-input single-output}
\acro{SMV}{single measurement vector}
\acro{SNR}[\textrm{SNR}]{signal-to-noise ratio} 
\acro{sp}[SP]{subspace pursuit}
\acro{SS}{spread spectrum}
\acro{SW}{sync word}
\acro{SAR}{Specific Absorption Rate}
\acro{SSB}{synchronization signal block}

\acro{TH}{time-hopping}
\acro{ToA}{time-of-arrival}
\acro{TR}{transmitted-reference}
\acro{TW}{Tracy-Widom}
\acro{TWDT}{TW Distribution Tail}
\acro{TCM}{trellis coded modulation}
\acro{TDD}{time-division duplexing}
\acro{TDMA}{Time Division Multiple Access}

\acro{UAV}{unmanned aerial vehicle}
\acro{uric}[URIC]{upper restricted isometry constant}
\acro{urict}[URICt]{upper restricted isometry constant threshold}
\acro{UWB}{ultrawide band}
\acro{UWBcap}[UWB]{Ultrawide band}   
\acro{URLLC}{Ultra Reliable Low Latency Communications}
         
\acro{wuric}[UWRIC]{upper weak restricted isometry constant}
\acro{wurict}[UWRICt]{UWRIC threshold}                
\acro{UE}{User Equipment}
\acro{UL}{uplink}
\acro{UB}{Upper Bound}

\acro{WiM}[WiM]{weigh-in-motion}
\acro{WLAN}{wireless local area network}
\acro{wm}[WM]{Wishart matrix}                               
\acroplural{wm}[WM]{Wishart matrices}
\acro{WMAN}{wireless metropolitan area network}
\acro{WPAN}{wireless personal area network}
\acro{wric}[WRIC]{weak restricted isometry constant}
\acro{wrict}[WRICt]{weak restricted isometry constant thresholds}
\acro{wrip}[WRIP]{weak restricted isometry property}
\acro{WSN}{wireless sensor network}                        
\acro{WSS}{wide-sense stationary}
\acro{WHO}{World Health Organization}
\acro{Wi-Fi}{wireless fidelity}

\acro{sss}[SpaSoSEnc]{sparse source syndrome encoding}

\acro{VLC}{visible light communication}
\acro{VPN}{virtual private network} 
\acro{RF}{Radio-Frequency}
\acro{FSO}{free space optics}
\acro{IoST}{Internet of space things}

\acro{GSM}{Global System for Mobile Communications}
\acro{2G}{second-generation cellular network}
\acro{3G}{third-generation cellular network}
\acro{4G}{fourth-generation cellular network}
\acro{5G}{5th-generation cellular network}	
\acro{gNB}{next-generation Node-B}
\acro{NR}{New Radio}
\acro{UMTS}{Universal Mobile Telecommunications Service}
\acro{LTE}{Long Term Evolution}

\acro{QoS}{Quality of Service}
\end{acronym}

\title{``Pencil Beamforming Increases Human Exposure to ElectroMagnetic Fields'': True or False?}
\author{Luca Chiaraviglio,$^{1,2}$ Simone Rossetti,$^{1,2}$ Sara Saida,$^{1,2}$ Stefania Bartoletti,$^{3}$ Nicola Blefari-Melazzi,$^{1,2}$\\
(1) Department of Electronic Engineering, University of Rome Tor Vergata, Rome, Italy, \\email luca.chiaraviglio@uniroma2.it\\
(2) CNIT, Italy, email \{simone.rossetti,sara.saida\}@cnit.it\\
(3) IEIIT-CNR, Bologna, Italy, email stefania.bartoletti@cnr.it\\
}

\maketitle

\IEEEpeerreviewmaketitle

\begin{abstract}
According to a very popular belief - very widespread among non-scientific communities - the exploitation of narrow beams, a.k.a. ``pencil beamforming'', results in a prompt increase of exposure levels radiated by 5G Base Stations (BSs). To face such concern with a scientific approach, in this work we propose a novel localization-enhanced pencil beamforming technique, in which the traffic beams are tuned in accordance with the uncertainty localization levels of User Equipment (UE). Compared to currently deployed beamforming techniques, which generally employ beams of fixed width, we exploit the localization functionality made available by the 5G architecture to synthesize the direction and the width of each pencil beam towards each served UE.  We then evaluate the effectiveness of pencil beamforming in terms of ElectroMagnetic Field (EMF) exposure and UE throughput levels over different realistic case-studies.
Results, obtained from a publicly released open-source simulator, dispel the myth: the adoption of localization-enhanced pencil beamforming triggers a prompt reduction of exposure w.r.t. other alternative techniques, which include e.g., beams of fixed width and cellular coverage not exploiting beamforming. The EMF reduction is achieved not only for the UE that are served by the pencil beams, but also over the whole territory (including the locations in proximity to the 5G BS). In addition, large throughput levels - adequate for most of 5G services - can be guaranteed when each UE is individually served by one dedicated  beam.

\end{abstract}

\begin{IEEEkeywords}
5G Cellular Networks, 5G Localization Service, Pencil Beam Management, EMF Analysis, Throughput Analysis
\end{IEEEkeywords}

\section{Introduction}
\label{sec:intro}

The deployment of 5G networks is a fundamental step to provide new services that are instrumental for many sectors, including e.g., e-health, smart transportation and industry 4.0. Although the service improvements triggered by 5G technology are clear and in general well recognized, the installation of 5G \acp{gNB} over the territory generates suspect and fear among part of the population (see e.g., \cite{beammyth,chiaraviglio2020health}), since the \ac{EMF} exposure radiated by 5G \acp{gNB} is believed to significantly increase compared to previous generations (e.g., 2G, 3G, 4G). Not surprisingly, anti-5G movements have gained attention in recent times, claiming that 5G constitutes a danger for human health and even inspiring sabotages to destroy masts/towers hosting 5G (and pre-5G) radio equipment \cite{5gfire}. However, the health risks allegations that are frequently associated with 5G exposure are not confirmed by scientific evidence, especially when the levels of exposure comply with international \ac{EMF} regulations \cite{bushberg2020ieee}. 

\begin{figure}[t]
\centering
\subfigure[No Beamforming]
{
	\includegraphics[width=4.1cm]{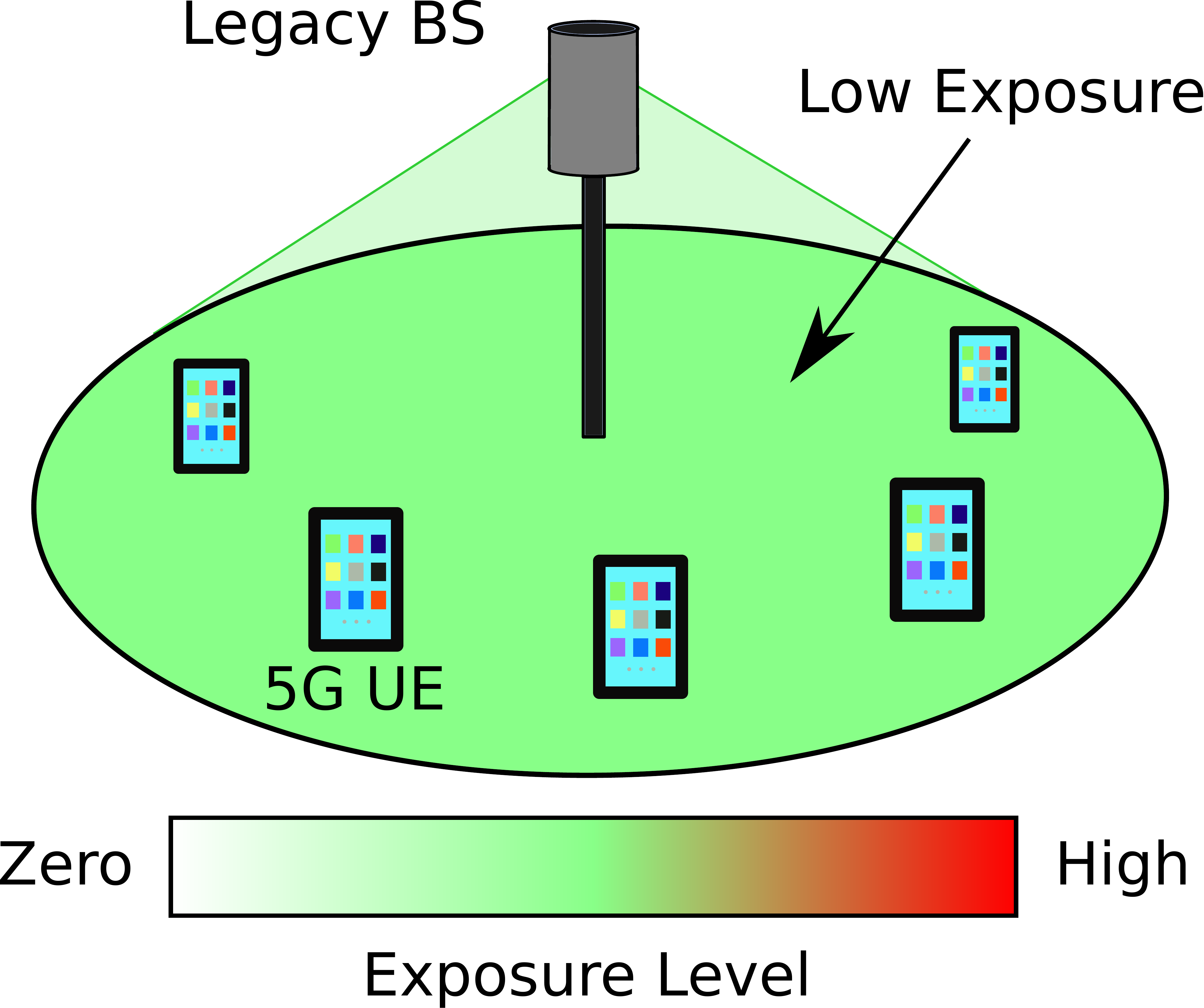}
	\label{fig:no_beamforming}
}
\subfigure[Pencil Beamforming]
{
	\includegraphics[width=4.1cm]{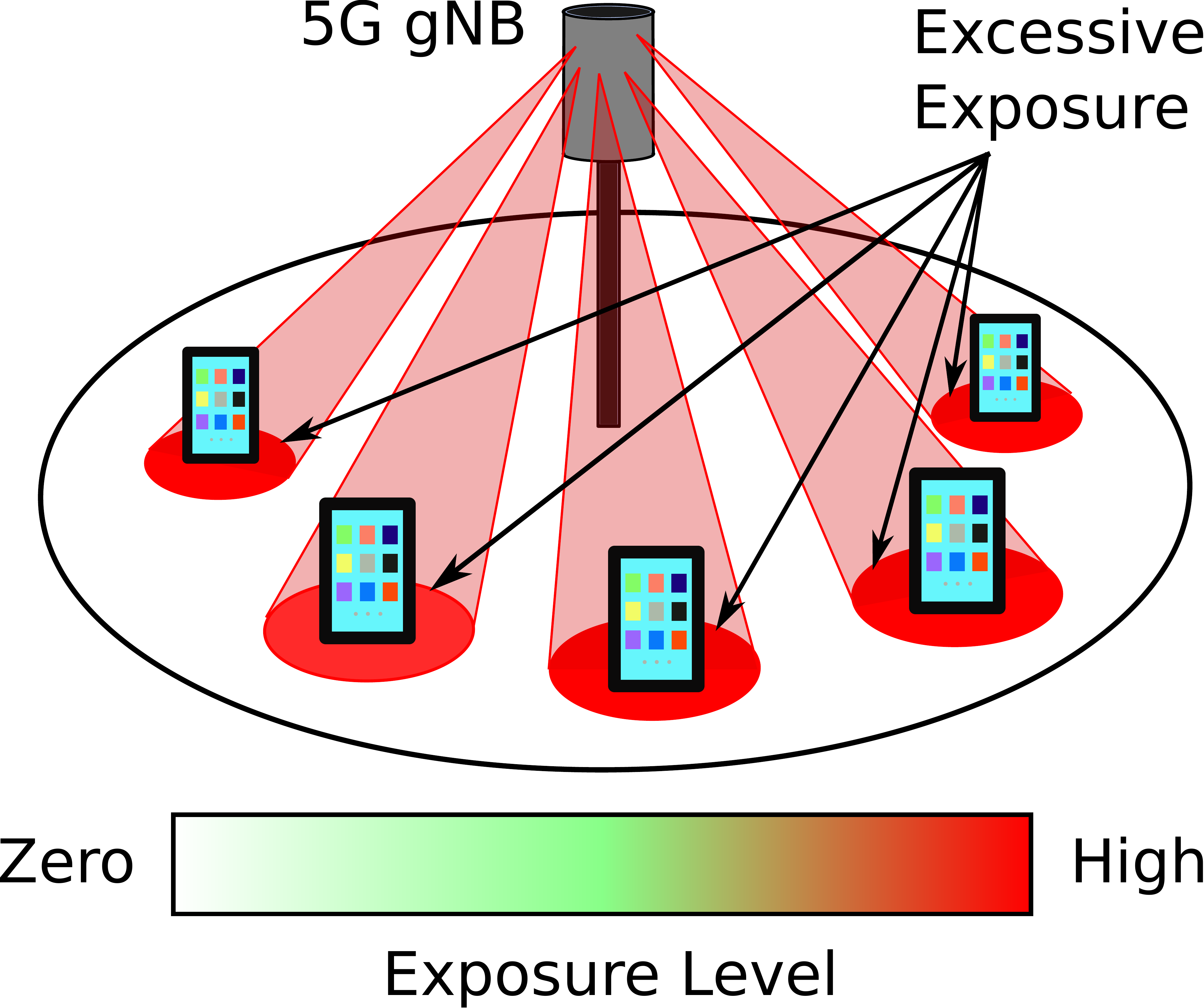}
	\label{fig:beamforming}
}
\caption{A popular layman belief: when \acp{gNB} exploiting pencil beamforming are employed (right), the level of exposure dramatically increases w.r.t. the case in which the cellular service is provided by BS not adopting beamforming (left). Is this allegation corroborated by scientific evidence or not? The goal of our work is to provide an answer to this intriguing question.}
\label{fig:myth_dep}
\end{figure}

In this context, the population's concerns against 5G frequently focus on the beamforming functionality \cite{beammyth,medicalpress2}. More specifically, a widespread opinion hypothesizes that the adoption of very narrow beams for serving the users (a.k.a. pencil beamforming) radically increases the \ac{EMF} levels radiated by 5G \acp{gNB} compared to wireless stations not implementing beamforming, thus posing a serious threat for the population, and in particular for those individuals who are radiated by the traffic beams. As sketched in Fig.~\ref{fig:no_beamforming}, the layman is firmly convinced that \acp{BS} not employing pencil beamforming radiate a pretty uniform and low exposure over the territory. On the other hand, the exploitation of 5G \acp{gNB} with pencil beamforming capabilities (Fig.~\ref{fig:beamforming}) is commonly associated with an excessive \ac{EMF} increase for the served users, thus fueling the population's concerns associated with 5G exposure. Despite the research community well knows that this is not the case, as the exposure levels from 5G \acp{gNB} always comply with \ac{EMF} regulations \cite{bushberg2020ieee} - hence ensuring health safety -, to the best of our knowledge, none of the previous works investigated the exposure of pencil beamforming (Fig.~\ref{fig:beamforming}), as well as its impact when compared to other solutions (e.g, no beamforming, like in Fig.~\ref{fig:no_beamforming}). Therefore, the hypothesis about an \ac{EMF} increase due to pencil beamforming is yet to be scientifically (and widely) refuted - even if the total exposure is still lower than the limit defined by laws.

More technically, the pencil beamforming functionality requires the localization of \ac{UE} that need to be served with the traffic beams. Intuitively, in fact, the knowledge of \ac{UE} positioning is essential to: \textit{i}) tune the pointing of the traffic beam(s) towards the served \ac{UE}, and \textit{ii}) adjust the width of the traffic beam in order to solely cover the area where the served \ac{UE} is localized (thus avoiding unwanted exposure/interference in the neighborhood of the beam). Clearly, both \textit{i}) and \textit{ii}) are two fundamental steps to increase the throughput levels and consequently to match the performance levels that are required by 5G services. In this scenario, the adoption of pencil beamforming requires the integration of \ac{UE} localization service in the 5G framework, a task that can be accomplished, e.g., by the \ac{LCS} functionality recently introduced by 3GPP in Rel. 16 \cite{3gpp.23.273}.





As a result, two fundamental questions emerge, namely: \textit{i}) Which is the impact of pencil beamforming on the \ac{EMF} exposure, especially in comparison to other solutions (e.g., no beamforming and/or traffic beams of fixed width)? \textit{ii}) How does the uncertainty level of the \ac{UE} location impact \ac{EMF} and throughput levels? The goal of this paper is to provide an answer to these intriguing questions, by tackling the problem in a way that can be understood even by non experts in the telecommunication field. More in depth, we design \textsc{5G-Pencil}, a framework for the scientific evaluation of localization-enhanced pencil beamforming. \textsc{5G-Pencil} implements a simple - yet effective - pencil beamforming policy that synthesizes the traffic beams by leveraging the 5G localization uncertainty level of each served user. We then code \textsc{5G-Pencil} in a publicly released open source simulator, in order to evaluate both the \ac{EMF} levels over the covered territory and the maximum downlink throughput achieved by each \ac{UE}.

Our results, obtained over a meaningful set of case studies with realistic parameters, scientifically confute the hypothesis that pencil beamforming increases the \ac{EMF} exposure. On the contrary, the localization-enhanced pencil beamforming guarantees a huge decrease of \ac{EMF} exposure, which is experienced not only by the served \ac{UE} but also over the whole covered territory. Moreover, the \ac{UE} downlink throughput matches the 5G requirements (especially the ones for \ac{eMBB} scenario), even when the \ac{UE} is served by one dedicated pencil beam. In addition, when the \ac{UE} location is precisely estimated (with an uncertainty localization level of few meters), very narrow and almost non overlapping pencil beams are synthesized by 5G \acp{gNB}, yielding to a general exposure reduction, which is also coupled by a substantial throughput increase. Eventually, we demonstrate that the widths of the synthesized pencil beams are within meaningful ranges for 5G radio equipment.

The rest of the paper is organized as follows. Sec.~\ref{sec:related_works} briefly reviews the related works. The main building blocks of \textsc{5G-Pencil} are reported in Sec.~\ref{sec:building_blocks}. Sec.~\ref{sec:beam_management} details the localization-enhanced pencil beamforming functionality of our framework. Sec.~\ref{sec:implementation} illustrates how  \textsc{5G-Pencil} is effectively implemented as an open source simulator. Results are reported in Sec.~\ref{sec:results}. Finally, Sec.~\ref{sec:conclusions} summarizes our work and highlights possible future avenues of research.

\section{Related Works}
\label{sec:related_works}
We compare our work w.r.t. the relevant literature, by considering the following taxonomy: \textit{i}) performance studies of beamforming in 5G networks, \textit{ii}) health risks associated with beamforming, and \textit{iii}) EMF assessment of beamforming.

\subsection{Performance of 5G Beamforming} 

The theoretical feasibility of beamforming as well as results from a prototype are presented in \cite{roh2014millimeter}. Yu \textit{et al.} \cite{yu2016load} propose a load balancing algorithm between macro cells and small cells, by extensively adopting a 3D beamforming feature. Awada \textit{et al.} \cite{awada2017simplified} design a channel model for signal measurement in a 5G cellular system that adopts narrow beams at fixed positions to serve the users. Ali \textit{et al.} \cite{ali2019system} derive closed-form expressions of the downlink \ac{SINR} levels when simultaneous beams are synthesized. In contrast to \cite{roh2014millimeter,yu2016load,awada2017simplified,ali2019system}, our work introduces a beam management policy that explicitly exploits a localization service to tune the pencil beams. In addition, we evaluate the impact of pencil beamforming not only on the throughput, but also on the \ac{EMF} exposure, which represents a major concern for the population. 

\subsection{Health Risks Associated with Beamforming} 

According to \cite{medicalpress2}, beamforming may pose an health risk, because 5G \acp{gNB} employing this feature will increase the effective radiated power. However, no scientific evidence is reported in order to support such claim. On the other hand, \cite{medicalpress} observes that beamforming allows to transmit signals only to the served users, while the other non-users will receive a lower amount of exposure. However, a rigorous evaluation of this effect is not included. \cite{NZbeam} reports that the adoption of beamforming may actually reduce the exposure compared to existing pre-5G technologies, without however providing any technical demonstration. Eventually, Bushberg \textit{et al.} \cite{bushberg2020ieee} observe that the zones that do not need to be served by active beams receive an \ac{EMF} exposure sharply lower w.r.t. the ones served by legacy equipment (e.g., 4G). 
In contrast to \cite{medicalpress2,medicalpress,NZbeam,bushberg2020ieee}, our work moves three steps further by: \textit{i}) designing a framework to technically analyze the pencil beamforming, \textit{ii}) introducing a pencil beamforming functionality that tunes the traffic beams in accordance with the \ac{UE} localization uncertainty level and \textit{iii}) evaluating the impact of pencil beamforming on \ac{EMF} and downlink throughput.

\subsection{EMF Assessment of Beamforming}

Thors \textit{et al.} \cite{Tho-17} evaluate realistic maximum power levels of a 5G \ac{gNB} employing beams focused on users. 
Nasim and Kim \cite{nasim2019adverse} evaluate the \ac{EMF} exposure from 5G \acp{gNB} employing beamforming. Their model, however, assumes only fixed (and wide) beams, thus neglecting the impact of pencil beams that are oriented (and tuned) towards the single users. Basikolo \textit{et al.} \cite{basikolo2019electromagnetic} show that the \ac{EMF} exposure from antennas employing beamforming complies with the limits defined by international regulations. Loh \textit{et al.} \cite{loh2020assessment} focus on the experimental and statistical assessment of the \ac{EMF} exposure from 5G \acp{gNB} in indoor environments, by considering beams oriented towards the \ac{UE}. Xu \textit{et al.} \cite{xu2020emf} show that simple models can be employed to determine the exclusion zones from \acp{gNB} operating with \ac{MIMO} and beamforming features. Adda \textit{et al.} \cite{adda2020theoretical} observe that only a fraction of the total \ac{gNB} power is radiated towards each single user, thus leading to a decrease of the \ac{EMF} levels. In contrast to \cite{Tho-17,nasim2019adverse,basikolo2019electromagnetic,loh2020assessment,xu2020emf,adda2020theoretical}, we introduce the following key novelties: \textit{i}) we design a framework that is able to synthesize the pencil beams in accordance with the \ac{UE} localization uncertainty level (a feature not exploited at all by previous works), and \textit{ii}) we demonstrate that the decrease of \ac{UE} localization uncertainty level can further reduce the \ac{EMF} generated by pencil beamforming.

\section{\textsc{5G-PENCIL} Framework: Building Blocks}
\label{sec:building_blocks}
\begin{figure}[t]
\centering
\includegraphics[width=\columnwidth]{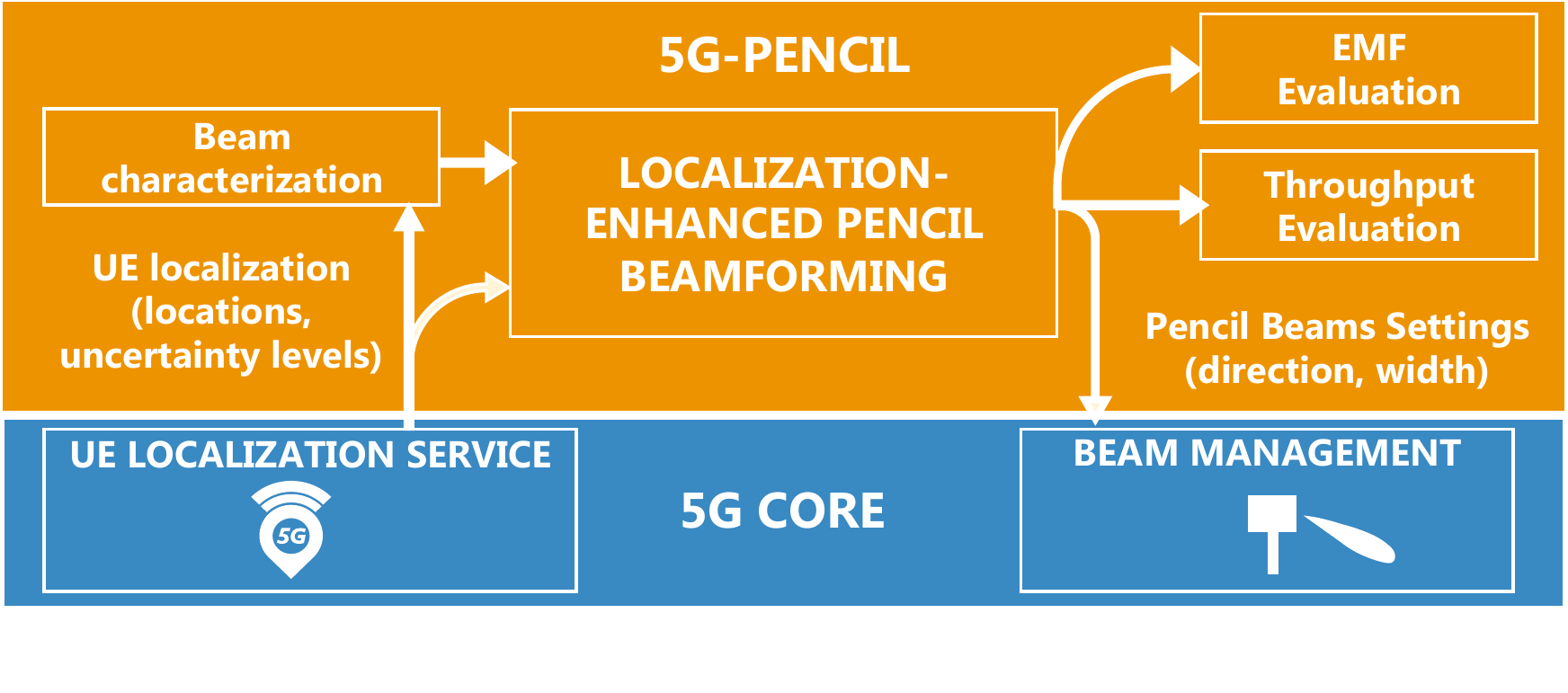}
\vspace{-1cm}
\caption{\textsc{5G-Pencil} framework: building blocks and interactions with the features implemented in the core of 5G architecture.}
\label{fig:building_blocks}
\end{figure}

Fig.~\ref{fig:building_blocks} reports the high-level view of the main building blocks that compose the \textsc{5G-Pencil} framework and their interaction with a set of core functionalities of the 5G architecture (i.e., external to our framework). More in depth,  
for each 5G user for which a localization service is available, an estimation of the \ac{UE} location and an associated uncertainty level is provided by the 5G core. Such information is then exploited by \textsc{5G-Pencil} to characterize the main features of each beam that needs to be synthesized, including e.g., selection of the coordinate system and computation of the pointing angles. The central part of the framework is the localization-enhanced pencil beamforming module, which computes the beam pointing and the beam width to serve each \ac{UE}, based on \ac{UE} localization as well as the information generated by the beam characterization module. This is a key innovative contribution of our work: when localization information is exploited to tune the beams, it is possible e.g., to synthesize narrow pencil beams towards the \ac{UE} that are localized with a given accuracy level. Moreover,  \textsc{5G-PENCIL} includes: \textit{i}) the \ac{EMF} evaluation module, which computes the exposure that is generated by the synthesized beams over the territory; and \textit{ii}) the throughput evaluation module to compute the downlink performance for each \ac{UE}. Finally, the pencil beam settings are passed to the 5G core for the beam management functionality, which implements the set of beams in the deployed 5G network. 



In the rest of the section, we provide more details about the following modules: \textit{a}) \ac{UE} localization, \textit{b}) beam characterization, \textit{c}) \ac{EMF} evaluation, \textit{d}) throughput evaluation and \textit{e}) beam management. We intentionally leave apart the innovative localization-enhanced pencil beamforming, which is detailed in Sec.~\ref{sec:beam_management}. In addition, we stress the fact that our paper is the first work to integrate together \textit{a})-\textit{e}) in an unique framework, which is instrumental for the localization-enhanced pencil beamforming and consequently for the goal of assessing the impact of pencil beamforming in terms of exposure (and throughput).

\subsection{User Equipment Localization}

\begin{figure}[t]
\centering
\includegraphics[width=7cm]{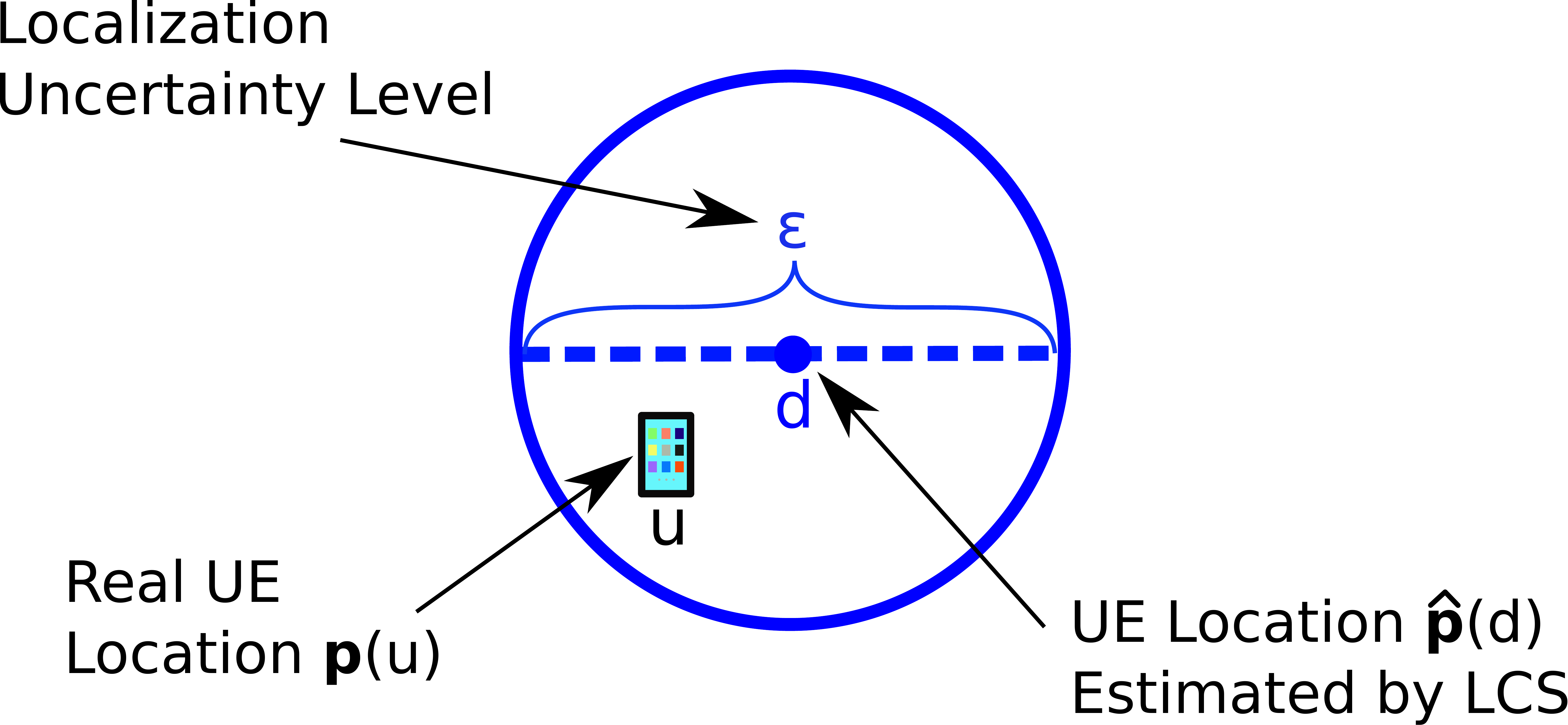}
\caption{Comparison among real \ac{UE} location $\mathbf{p}(u)$ and \ac{UE} location $\hat{\mathbf{p}}(d)$ estimated by \ac{LCS}. The figure reports also the uncertainty level for the estimated \ac{UE} location as a circle of diameter $\epsilon$.}
\label{fig:circle}
\end{figure}

5G \ac{UE} localization  is  introduced  in  Rel. 15 of 3GPP, through the definition of the location management function within 5G positioning  \cite{3gpp.23.273}. The UE location is computed from  measurements mainly based on \ac{DL-TDoA} and beamforming \ac{AoA} between the \acp{gNB} and the \ac{UE}. Position accuracy is defined by 3GPP as the difference between actual location and estimated location and it is thus related to the uncertainty level of the \ac{UE} position. More specifically, the \ac{UE} location uncertainty level  varies across the network geographic area depending on the \ac{UE} true position, due to a multitude of factors, which include, e.g., variability of radio conditions, cell configuration and cell density. In addition, the accuracy can be negotiated based on the requirements of the specific 5G service that needs to be provided. According to the 3GPP technical specifications in Rel. 16 \cite{3GPP:TS:22.071:V16.0.0}, the \ac{LCS} shall satisfy or approach as closely as possible the requested or negotiated accuracy when other \ac{QoS} parameters are not in conflict.

More formally, the achieved accuracy level of location information is expressed through the shapes and the uncertainty areas defined in  \cite{3GPP:TS:23.032:V16.0.0}. An uncertainty circle indicates a point when its position is known only with a limited accuracy, with an uncertainty level that is described by the circle diameter.\footnote{Alternatively, the uncertainty level can be expressed as the circle radius. In this work, however, we stick to the assumption that the uncertainty level corresponds to the circle diameter.} To this aim,  Fig.~\ref{fig:circle} reports a representative example in which $\hat{\mathbf{p}}(d)$ denotes the estimated \ac{UE} position, which is defined as an integer index $d$ to extract a row of x-y-z coordinates from array $\hat{\mathbf{p}}$. The circle of diameter $\epsilon$ is the localization uncertainty area in the horizontal plane. Finally, $\mathbf{p}(u)$ is the real \ac{UE} location, which is expressed as integer index $u$ to extract a row of x-y-z coordinates from array $\mathbf{p}$. For the sake of simplicity, we get rid of coordinates arrays $\mathbf{p}$ and $\hat{\mathbf{p}}$ from now on, by simply referencing to the real (estimated) \ac{UE} location through index $u$ ($d$).

In our context, the \ac{UE} localization is provided by the \ac{LCS} functionality implemented in the 5G network. As a result, \textsc{5G-Pencil} takes as input the estimated location $d$ for each \ac{UE}, together with the uncertainty level $\epsilon$.

\begin{figure}[t]
\centering
\subfigure[Azimuth]
{
	\includegraphics[width=4.9cm]{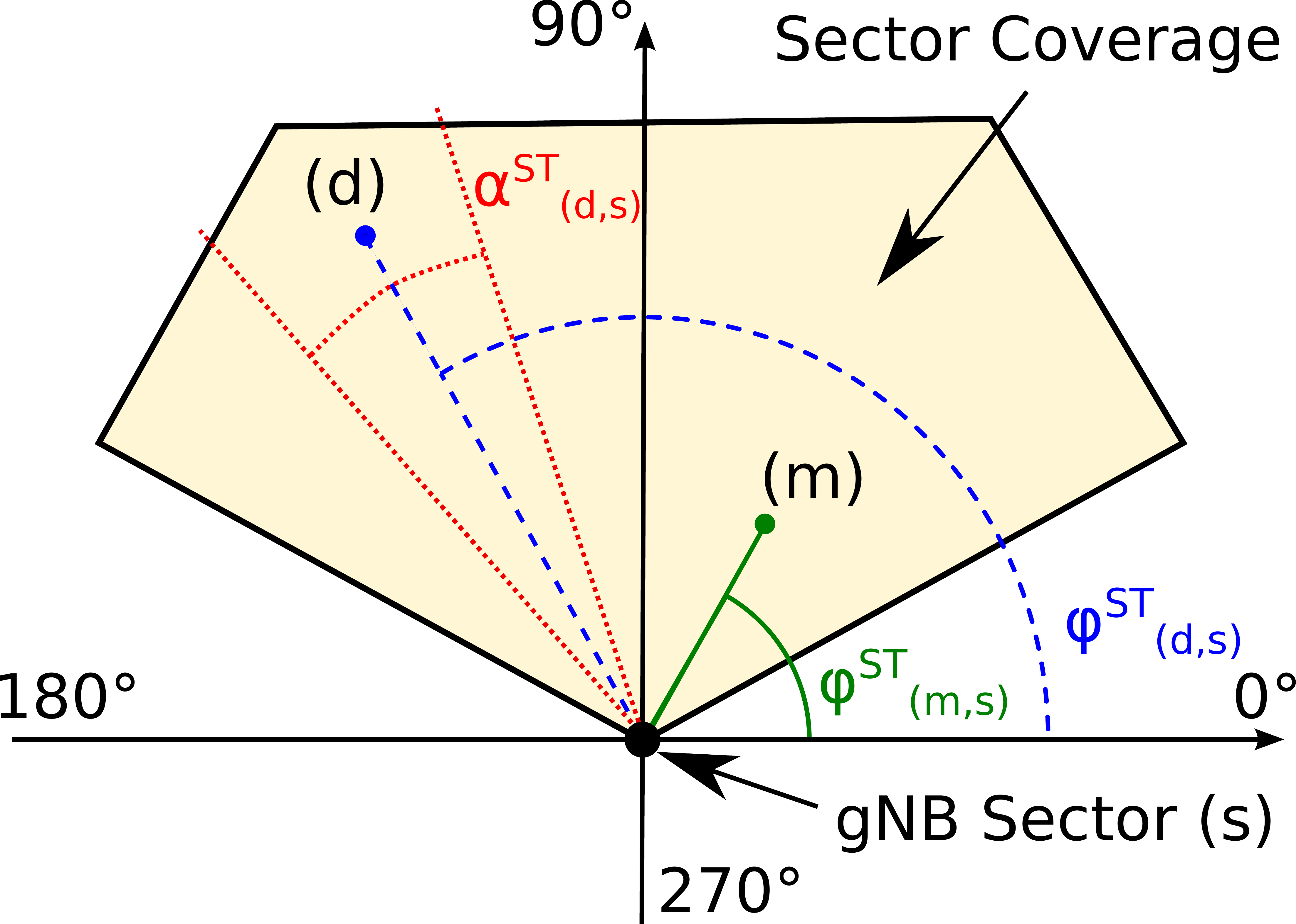}
	\label{fig:angles_steering}
}
\subfigure[Elevation]
{
	\includegraphics[width=3.3cm]{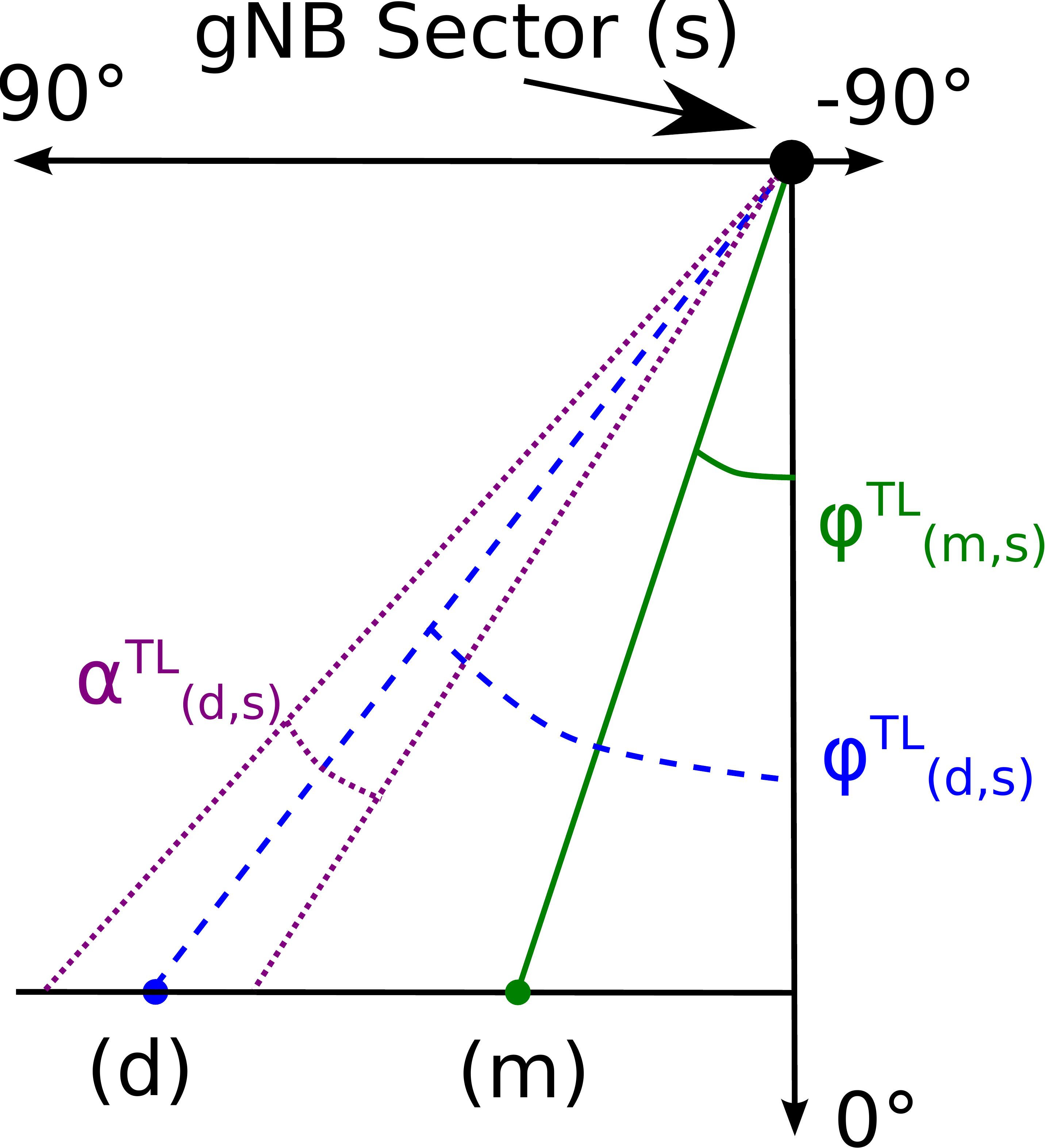}
	\label{fig:angles_tilting}
}
\caption{Definition of the angles for a toy case scenario with a sector coverage equal to 1/3 of an hexagon, one beam deployed over $d$ from $s$ and one measurement spot $m$. }
\label{fig:angles_beam}
\vspace{-4mm}
\end{figure}

\subsection{Beam Characterization}
\label{subsec:beam}

Let us consider a generic sector $s$ of 5G \ac{gNB} and two points $u$  and $m$  located in the territory. Point $d$ is the estimated \ac{UE} location already introduced in Fig.~\ref{fig:circle}. Such position is assumed to be the target of a deployed beam, and therefore it is referred hereafter as ``deployment spot''. On the other hand, $m$  is named ``measurement spot'', because such location is used to evaluate \ac{EMF} that is generated by the beam targeting $d$. 
Let us then assume two observation planes for the sector, one horizontal (i.e., azimuth) and one vertical (i.e., elevation). Focusing on the horizontal plane, we introduce an angular orientation system centered on $s$, spanning from 0$^{\circ}$ to 360$^{\circ}$, as shown in Fig.~\ref{fig:angles_steering}. Let us denote with $\phi^{\text{ST}}_{(d,s)}$ the steering angle of $d$ w.r.t. sector $s$. In a similar way, $\phi^{\text{ST}}_{(m,s)}$ is the steering angle that is measured over $m$ from $s$. Focusing then on the vertical plane, we introduce an angular orientation system between 90$^{\circ}$ and -90$^{\circ}$, again centered on $s$, as shown in Fig.~\ref{fig:angles_tilting}. We then denote as $\phi^{\text{TL}}_{(d,s)}$ and $\phi^{\text{TL}}_{(m,s)}$ the tilting angles of $d$ and $m$ w.r.t. sector $s$ respectively.  

In the following, we introduce the notation to characterize the beam widths. Specifically, we adopt the commonly-used assumption that the width is denoted by the cone where the beam gain is at most 3~[dB] lower than the maximum value (i.e., the one achieved over $\phi^{\text{ST}}_{(d,s)}$ and $\phi^{\text{TL}}_{(d,s)}$). More in depth, let us denote with $\alpha^{\text{ST}}_{(d,s)}$ the projection on the horizontal plane of the 3~[dB] beam cone used to serve deployment spot $d$ from $s$. Similarly, let us denote with $\alpha^{\text{TL}}_{(d,s)}$ the projection of the same 3~[dB] beam cone on the vertical plane. As shown in Fig.~\ref{fig:angles_beam}, both $\alpha^{\text{ST}}_{(d,s)}$ and $\alpha^{\text{TL}}_{(d,s)}$ are expressed as relative angles, and therefore they do not depend on the absolute angular positioning systems adopted for $s$. Intuitively, the setting of $\alpha^{\text{ST}}_{(d,s)}$ and $\alpha^{\text{TL}}_{(d,s)}$ heavily influences the size of the area covered by the beam, which in turns affects both throughput and \ac{EMF} values. Consequently, the optimization of $\alpha^{\text{ST}}_{(d,s)}$ and $\alpha^{\text{TL}}_{(d,s)}$ values is one of the key goals that are targeted by our framework.

To summarize, the beam that covers deployment spot $d$ from $s$ is fully denoted by the quadruple $\phi^{\text{ST}}_{(d,s)}$, $\phi^{\text{TL}}_{(d,s)}$, $\alpha^{\text{ST}}_{(d,s)}$, $\alpha^{\text{TL}}_{(d,s)}$. In addition, each measurement spot $m$ is characterized by angles $\phi^{\text{ST}}_{(m,s)}$ and $\phi^{\text{TL}}_{(m,s)}$. 

\subsection{EMF Evaluation}
\label{subsec:emf_eval}


Let us assume that a pencil beam is deployed over $d$ by a radiating element installed on $s$, and that we want to compute the exposure generated by this beam over $m$.  In line with \ac{ITU} recommendations \cite{iturec70,iturec91}, as well as previous works \cite{Tornevik}, we assume an exclusion zone - whose access is prohibited to the general public - in proximity to the \ac{gNB}. Since our goal is to evaluate the impact of pencil beamforming on the population (i.e., not for maintenance workers who may sporadically operate inside the \ac{gNB} exclusion zone), both $m$ and $d$ are outside the exclusion zone of $s$. Consequently, the \ac{EMF} is always evaluated in the far-field region, where the exposure is expressed in terms of \ac{EMF} strength and/or \ac{PD} \cite{iturec70,iturec91}.\footnote{An alternative metric to characterize the exposure is the \ac{SAR}, which is usually employed to evaluate the exposure levels in near-field regions, especially for personal devices operating in close proximity to the body, like smartphones. Since our goal is to evaluate the level of exposure from \acp{gNB} in the far-field region,  the evaluation of exposure in terms of \ac{EMF} strength / \ac{PD} provides a more meaningful information than \ac{SAR}.} This choice is also supported by the fact that the exposure limits from \acp{gNB} are normally defined as maximum \ac{EMF} strength (or, equivalently, as maximum \ac{PD}) \cite{international2020guidelines}. In addition, the on-the-field measurement of the exposure levels from \acp{gNB} is typically performed by employing meters that  measure \ac{EMF} strength \cite{iturec91}.

More formally, we start from the widely-accepted point-source model of \ac{ITU} \cite{iturec70} to compute the \ac{PD} $S_{(d,s,m)}$ that is received by $m$ from the pencil beam serving $d$ from $s$:\footnote{The suitability of using the point-source model for evaluating the exposure of \acp{gNB} employing beamforming is also confirmed by \cite{xu2020emf}.}
\begin{equation}
\label{eq:point_source}
S_{(d,s,m)} = \frac{P^{\text{EIRP}}_{s}\cdot F_{(d,s,m)}}{4 \pi \cdot \delta_{(s,m)}^2}
\end{equation}
where $P^{\text{EIRP}}_{s}$ is the \ac{EIRP} of the antenna element installed on sector $s$, $F_{(d,s,m)} \in (0,1]$ is the antenna numeric gain observed over $m$ w.r.t. the pencil beam serving $d$ from $s$, and $\delta_{(s,m)}$ is the 3D distance between $s$ and $m$.

The term $P^{\text{EIRP}}_{s}$ is then expressed as:
\begin{equation}
\label{eq:eirp}
P^{\text{EIRP}}_s = P^{\text{MAX}}_{s} \cdot G^{\text{MAX}} 
\end{equation}
where $P^{\text{MAX}}_{s}$ is the maximum radiated power by one antenna element on sector $s$ and $G^{\text{MAX}}$ is the maximum antenna gain.

Focusing instead on $F_{(d,s,m)}$, we express the normalized numeric gain as in \cite{iturec70}:
\begin{equation}
\label{eq:num_gain}
F_{(d,s,m)}=\left(10^{\frac{A^{\text{AZ}}_{(d,s,m)}+A^{\text{EL}}_{(d,s,m)}}{10}}\right)^2 
\end{equation}
where $A^{\text{AZ}}_{(d,s,m)}$ and $A^{\text{EL}}_{(d,s,m)}$ are the azimuth and the elevation radiation patterns (in [dB]) observed over $m$ from an antenna element in sector $s$ serving deployment spot $d$.

Both $A^{\text{AZ}}_{(d,s,m)}$ and $A^{\text{EL}}_{(d,s,m)}$ are then expressed as in \cite{yu2016load}, by adopting the angles already introduced to characterize the beam over $d$ and the measurement spot $m$:
\begin{equation}
\label{eq:az}
A^{\text{AZ}}_{(d,s,m)}=-\text{min}\left[12\left(\frac{\phi^{\text{ST}}_{(m,s)}-\phi^{\text{ST}}_{(d,s)}}{\alpha^{\text{ST}}_{(d,s)}}\right)^2,A^{\text{AZ}}_{\text{MIN}}\right]
\end{equation}
\begin{equation}
\label{eq:al}
A^{\text{EL}}_{(d,s,m)}=-\text{min}\left[12\left(\frac{\phi^{\text{TL}}_{(m,s)}-\phi^{\text{TL}}_{(d,s)}}{\alpha^{\text{TL}}_{(d,s)}}\right)^2,A^{\text{EL}}_{\text{MIN}}\right]
\end{equation}
where $A^{\text{AZ}}_{\text{MIN}}$ and $A^{\text{EL}}_{\text{MIN}}$ are the front-to-back ratio and the side lobe level limit, respectively. By analyzing in detail Eq.~(\ref{eq:az}), we can note that the maximum radiation patterns are achieved when $\phi^{\text{ST}}_{(m,s)}=\phi^{\text{ST}}_{(d,s)}$ and $\phi^{\text{TL}}_{(m,s)}=\phi^{\text{TL}}_{(d,s)}$, i.e., the measurement spot $m$ is co-located with $d$. In addition, the beam widths $\alpha^{\text{ST}}_s$ and $\alpha^{\text{TL}}_s$ act as scaling parameters for the radiation pattern: the higher is the beam width, the lower is the impact of the steering/tilting angles and consequently the larger is the radiation pattern. Finally, the front-to-back ratio and the side lobe level limit are used to bound the minimum radiation pattern values.

In the following, we extend the exposure model by considering a set of deployment spots $\mathcal{D}$ and a set of \ac{gNB} sectors $\mathcal{S}$. To this aim, let us introduce parameter $X_{(d,s)}$, taking value 1 if $d \in \mathcal{D}$ is served by $s \in \mathcal{S}$, 0 otherwise. 
The overall exposure over measurement point $m$ by all the beams that are deployed in the scenario is then expressed as:
\begin{equation}
\label{eq:tot_s}
S^{\text{TOT}}_{m}=\sum_{s \in \mathcal{S}} \sum_{d \in \mathcal{D}} S_{(d,s,m)} \cdot X_{(d,s)}
\end{equation}

Finally, we exploit the widely-known equivalence between \ac{EMF} strength and \ac{PD} \cite{iturec70} (which we remind is valid in the far-field region) to compute the total \ac{EMF} strength observed in $m$:
\begin{equation}
\label{eq:tot_e}
E^{\text{TOT}}_{m}=\sqrt{S^{\text{TOT}}_{m} \cdot Z} 
\end{equation}
where $Z=377$~[$\Omega$] is the free-space wave impedance.

Two considerations hold by analyzing Eq.~(\ref{eq:point_source})-(\ref{eq:tot_e}). First, the total exposure generated by multiple beams deployed in the scenario is evaluated for each measurement spot $m \in \mathcal{M}$. Second, the point-source model of Eq.~(\ref{eq:point_source}) is an upper bound of the actual level of exposure that is measured on-the-field \cite{iturec70}, thus substantiating the outcomes of our work.

\subsection{Throughput Evaluation}
\label{subsec:thr_eval}

We initially evaluate the throughput that is received by a generic user located at position $u$ inside deployment spot $d$. For simplicity, we assume that deployment spot $d$ is served by one single dedicated beam that is radiated by sector $s$. The maximum downlink throughput $T_{(d,s,u)}$ is computed with the classical Shannon capacity model:
\begin{equation}
\label{eq:tp}
T_{(d,s,u)} =  B_{s} \cdot \log_{2}(1 + \text{SINR}_{(d,s,u)})
\end{equation}
where $B_{s}$ is the adopted bandwidth and $\text{SINR}_{(d,s,u)}$ is the \ac{SINR} experienced at $u$, due to the beam that is deployed over $d$ from $s$.

We then express the \ac{SINR} as the one used by the multiple-beam system \cite{ali2019system}:
\begin{equation}
\label{eq:sinr}
\text{SINR}_{(d,s,u)}=\frac{P^{\text{RX}}_{(d,s,u)}}{\underbrace{\sum_{d'\neq d} P^{\text{RX}}_{(d',s,u)}}_{\text{Intra-sector Interference}} + \underbrace{\sum_{d'\neq d} \sum_{s' \neq s}  P^{\text{RX}}_{(d',s',u)}}_{\text{Inter-sector Interference}} + N}    
\end{equation}
where $P^{\text{RX}}_{(d,s,u)}$ is the power received at $u$ from the beam deployed over $d$ by $s$ and $N$ is the noise component. From Eq.~(\ref{eq:sinr}), we can note that each beam can interfere with all the others that are deployed from the same sector (intra-sector term) and/or from other sectors (inter-sector term). In order to evaluate the throughput under different interference assumptions, in this work we consider the computation of the \ac{SINR} with and without intra-sector interference. Clearly, when the intra-sector term is not considered, the throughput tends to be higher, because the beams generated by the same sector do not interfere with each other.

In line with \cite{ali2019system}, we express the received power $P^{\text{RX}}_{(d,s,u)}$ (in [dB]) as:
\begin{equation}
\begin{aligned}
\label{eq:Pucb}
P^{\text{RX}}_{(d,s,u)}  = \underbrace{P^{\text{TX}}_{s}}_{\text{Max. Tx Power}} - \underbrace{L^{\text{PL}}_{(s,u)}}_{\text{3D Path Loss}} + \underbrace{A^{\text{AZ}}_{(d,s,u)} + A^{\text{EL}}_{(d,s,u)}}_{\text{Beam radiation pattern}} + \\ 
\underbrace{G^{\text{TX}}_s}_{\text{Max. Tx Gain}} + \underbrace{B^{\text{AZ}}_{(d,s,u)} + B^{\text{EL}}_{(d,s,u)} + G^{\text{BF}}_s}_{\text{Beamforming gain}} 
\end{aligned}
\end{equation}
where $P^{\text{TX}}_{s}$ is the maximum transmission power of the entire antenna array located at $s$, $L^{\text{PL}}_{(s,u)}$ is the 3D path loss term between $s$ and $u$, $A^{\text{AZ}}_{(d,s,u)}$ and $A^{\text{EL}}_{(d,s,u)}$ are the antenna radiation patterns already defined in Eq.~(\ref{eq:az}),(\ref{eq:al}) (computed here w.r.t. \ac{UE} location $u$), $G^{\text{TX}}_s$ is the maximum transmission gain of one antenna element in $s$, $G^{\text{BF}}_s$ is the maximum beamforming gain for $s$, while the beamforming terms $B^{\text{AZ}}_{(d,s,u)}$ and $B^{\text{EL}}_{(d,s,u)}$ are formally expressed as in \cite{awada2017simplified}:
\begin{equation}
\label{eq:baz}
B^{\text{AZ}}_{(d,s,u)}=10\log_{10}\left[\mathrm{sinc}\left(\frac{\phi^{\text{ST}}_{(u,s)}-\phi^{\text{ST}}_{(d,s)}}{1.13 \cdot \alpha^{\text{ST}}_{(d,s)}}\right)^{2}\right]
\end{equation}
\begin{equation}
\label{eq:bal}
B^{\text{EL}}_{(d,s,u)}=10\log_{10}\left[\mathrm{sinc}\left(\frac{\phi^{\text{TL}}_{(u,s)}-\phi^{\text{TL}}_{(d,s)}}{1.13 \cdot \alpha^{\text{TL}}_{(d,s)}}\right)^{2}\right]
\end{equation}

By analyzing in detail Eq.~(\ref{eq:Pucb})-(\ref{eq:bal}), we can note that the received power is computed as a combination of different terms that scale the maximum transmission power of the entire antenna array. As a result, the level of detail provided by the throughput model is higher compared to the \ac{EMF} one. However, also in this case the beam widths severely impact both the beam radiation patterns and the beamforming gain terms, thus affecting the received power and hence in turn the received throughput.

\subsection{Beam Management}
\label{subsec:beam_management_functionality}

The set of traffic beams that are selected by our framework is then passed as input to the beam management module of the 5G network. Apart from synthesizing the pencil beams on the deployed \acp{gNB}, this module controls the set of beam(s) that are deployed to provide basic coverage (a.k.a. ``broadcast beams'') and/or for retrieving the \ac{UE} localization information. However, the exposure from such additional beams is overall much lower than the \ac{EMF} radiated by the traffic beams, as shown e.g., in \cite{adda2020theoretical}. Therefore, the \ac{EMF} evaluation in our framework is intentionally focused on the impact of traffic beams in terms of \ac{EMF} (and throughput).



\begin{algorithm}[t]
	\caption{Localization-Enhanced Pencil Beam Tuning}\label{alg:pseudocode}
	\begin{algorithmic}[1]
		\State \textbf{Input:} $\mathcal{S}$, $\mathcal{D}$, $\epsilon$, $\alpha^{\text{ST}}_{\text{MIN}}$, $\alpha^{\text{TL}}_{\text{MIN}}$
		\State \textbf{Output:} $X_{(d,s)}$, $\alpha^{\text{ST}}_{(d,s)}$, $\alpha^{\text{TL}}_{(d,s)}$
		\For{$s$ in $\mathcal{S}$}
			\For{$d$ in $\mathcal{D}$}
			    \If{check\_coverage($d$,$s$) == true}
			        \State \textit{// Deployment spot - sector association}
			        \State $X_{(d,s)}=1$; 
			        \State \textit{// Interception points computation}
			        \State [$I^{\text{W}}_{(d,s)}$,$I^{\text{E}}_{(d,s)}$,$I^{\text{N}}_{(d,s)}$,$I^{\text{S}}_{( d ,s)}$] = comp\_pt($d$,$s$,$\epsilon$);
			        \State \textit{//Horizontal width setting}
			        \State $\lambda^{\text{H}}_c$ = $I^{\text{W}}_{(d,s)} \xrightarrow{\text{H}} (s)$;
			        \State $\lambda^{\text{H}}_b$ = $\lambda^{\text{H}}_a$;
			        \State $\lambda^{\text{H}}_a$ = $I^{\text{W}}_{(d,s)} \xrightarrow{\text{H}} I^{E}_{(d,s)}$;
			            \State  $\text{tmp\_angle\_st}=\arccos{\left(1-\frac{(\lambda^{\text{H}}_a)^2}{2 \cdot (\lambda^{\text{H}}_c)^2}\right)}$; 
			        \State $\alpha^{\text{ST}}_{(d,s)}=\max(\text{tmp\_angle\_st},\alpha^{\text{ST}}_{\text{MIN}})$;
			        \State \textit{//Vertical width setting}
			        \State $\lambda^{\text{V}}_c$ = $I^{\text{N}}_{(d,s)} \xrightarrow{\text{V}} (s)$;
			        \State $\lambda^{\text{V}}_b$ = $I^{\text{S}}_{(d,s)} \xrightarrow{\text{V}} (s)$;
			        \State $\text{tmp\_angle\_tl}=\arccos{\left(\frac{(\lambda^{\text{V}}_c)^2+(\lambda^{\text{V}}_b)^2-\epsilon^2}{2 \cdot \lambda^{\text{V}}_c \cdot \lambda^{\text{V}}_b}\right)}$; 
			        \State $\alpha^{\text{TL}}_{(d,s)}=\max(\text{tmp\_angle\_tl},\alpha^{\text{TL}}_{\text{MIN}})$;
				\EndIf
			\EndFor
		\EndFor
	\end{algorithmic}
\end{algorithm}

\section{Localization-Enhanced Pencil Beamforming}
\label{sec:beam_management}

We initially introduce a set of simplifying and/or conservative assumptions that are instrumental for the pencil beamforming module. Then, we detail the algorithm that we have designed to tune the pencil beams on the deployment spots.

\subsection{Main Assumptions}

We consider the adoption of antenna arrays composed of a large number of radiating elements, denoted as $N^\text{R}_s$. Each \ac{gNB} hosts a set of non-overlapping sectors, each of them equipped with an antenna array. Each radiating element is able to generate a traffic beam that is characterized by a given direction and by a given width in the area covered by the sector. We then assume that each deployment spot $d \in \mathcal{D}$ is served by at most one dedicated pencil beam. Although this assumption may appear relatively conservative at a first glance, as $d$ may be alternatively served by multiple pencil beams at the same time, we demonstrate that the throughput level achieved with one pencil beam over $d$ already matches the performance level required by 5G, thus substantiating our analysis.\footnote{The investigation of multiple pencil beams serving each \ac{UE} is left for future work.} Moreover, we assume that the number of radiating elements $N^\text{R}_s$ is always higher or equal than the number of deployment spots placed in each sector. In this way, all the \ac{UE} in the territory are served by our framework. Moreover, we assume that: \textit{i)} the pencil beams are activated all together at the same time, and  \textit{ii)}  each pencil beam always transmits at the maximum power in the downlink direction, i.e., no traffic adaptation mechanisms are applied to the power radiated by the beam. In this way, we evaluate \ac{EMF} (throughput) under high exposure (peak traffic) conditions. 

\begin{figure}[t]
\centering
\includegraphics[width=6cm]{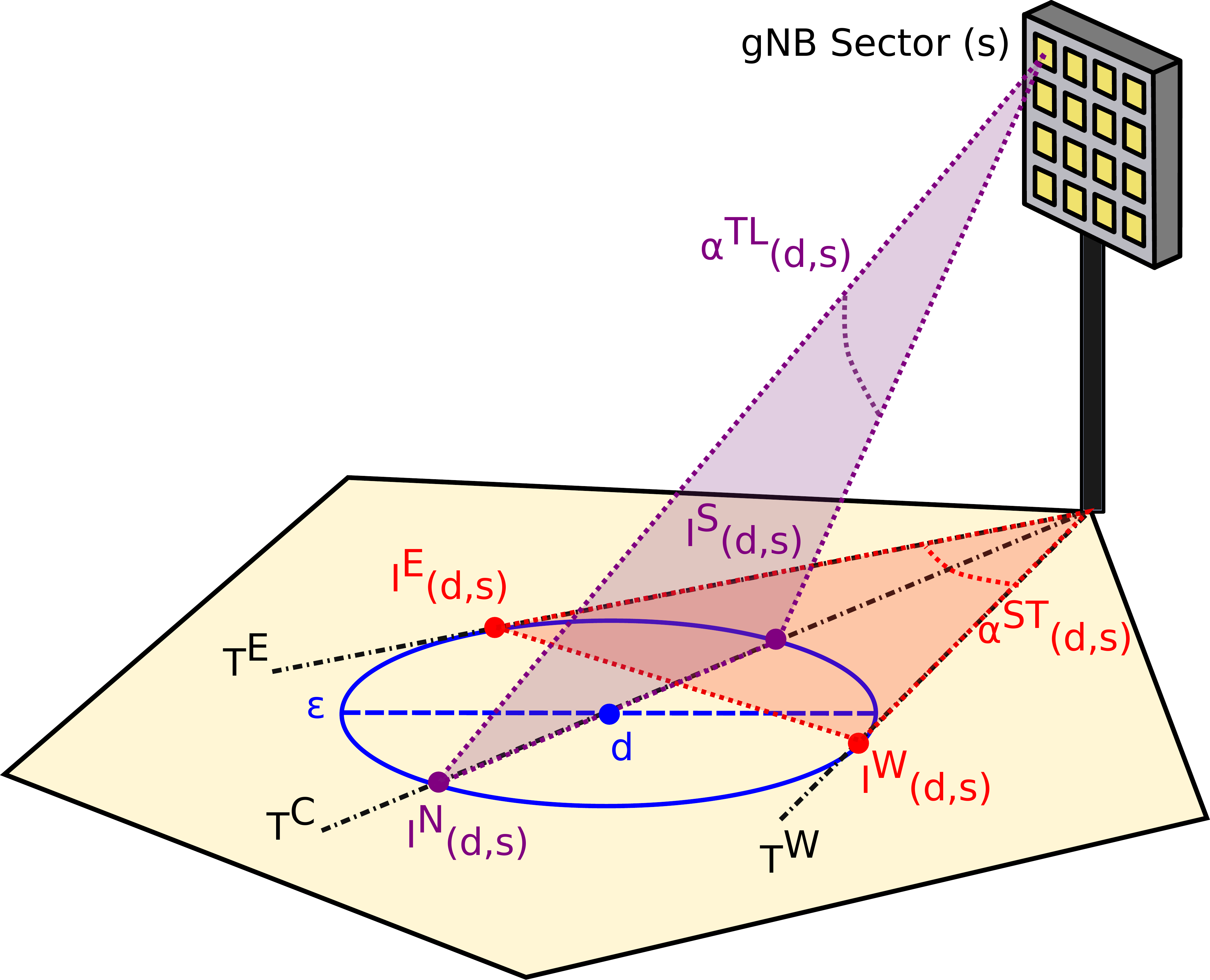}
\caption{Graphical sketch of the interception points and the triangles that are used to tune the pencil width.}
\label{fig:beam_aplitudes}
\end{figure}

\subsection{Algorithm Description}

Alg.~\ref{alg:pseudocode} reports the high-level pseudocode of the pencil beam setting procedure implemented in \textsc{5G-Pencil}. The algorithm requires as input the set of sectors $\mathcal{S}$, the set of deployment spots $\mathcal{D}$ and the localization uncertainty level $\epsilon$ (which are passed to \textsc{5G-Pencil} by the \ac{UE} localization service module), as well as the minimum values for the beam width in the horizontal and vertical planes, denoted as $\alpha^{\text{ST}}_{\text{MIN}}$ and $\alpha^{\text{TL}}_{\text{MIN}}$, respectively. Intuitively, in fact, the minimum beam widths are constrained by the technological features of the antenna arrays, which can synthesize pencil beam of a given width up to minimum values $\alpha^{\text{ST}}_{\text{MIN}}$ and $\alpha^{\text{TL}}_{\text{MIN}}$. The algorithm then returns as output the deployment spot-sector association array $X_{(d,s)}$, as well as the selected pencil width settings $\alpha^{\text{ST}}_{(d,s)}$ and $\alpha^{\text{TL}}_{(d,s)}$.

Initially, (lines 3-4), the algorithm iterates over the elements in $\mathcal{S}$ and in $\mathcal{D}$. For each pair $(d,s)$, a coverage check is performed. This function is intentionally not expanded Alg.~\ref{alg:pseudocode}, due to the fact that the $(d,s)$ association is a choice left to the operator, which may associate a deployment spot to a sector depending on multiple factors, including, e.g., link budget evaluation at the \ac{UE} location, maximum coverage distance between $s$ and $d$, traffic load distribution among the \acp{gNB} and/or a mixture between them.  If $d$ can be covered by $s$, then the serving variable $X_{(d,s)}$ is set to 1 (line 7) and a pencil beam is tuned on $d$ from $s$ (lines 8-20). 

\begin{figure*}[t]
\centering
\includegraphics[width=16cm]{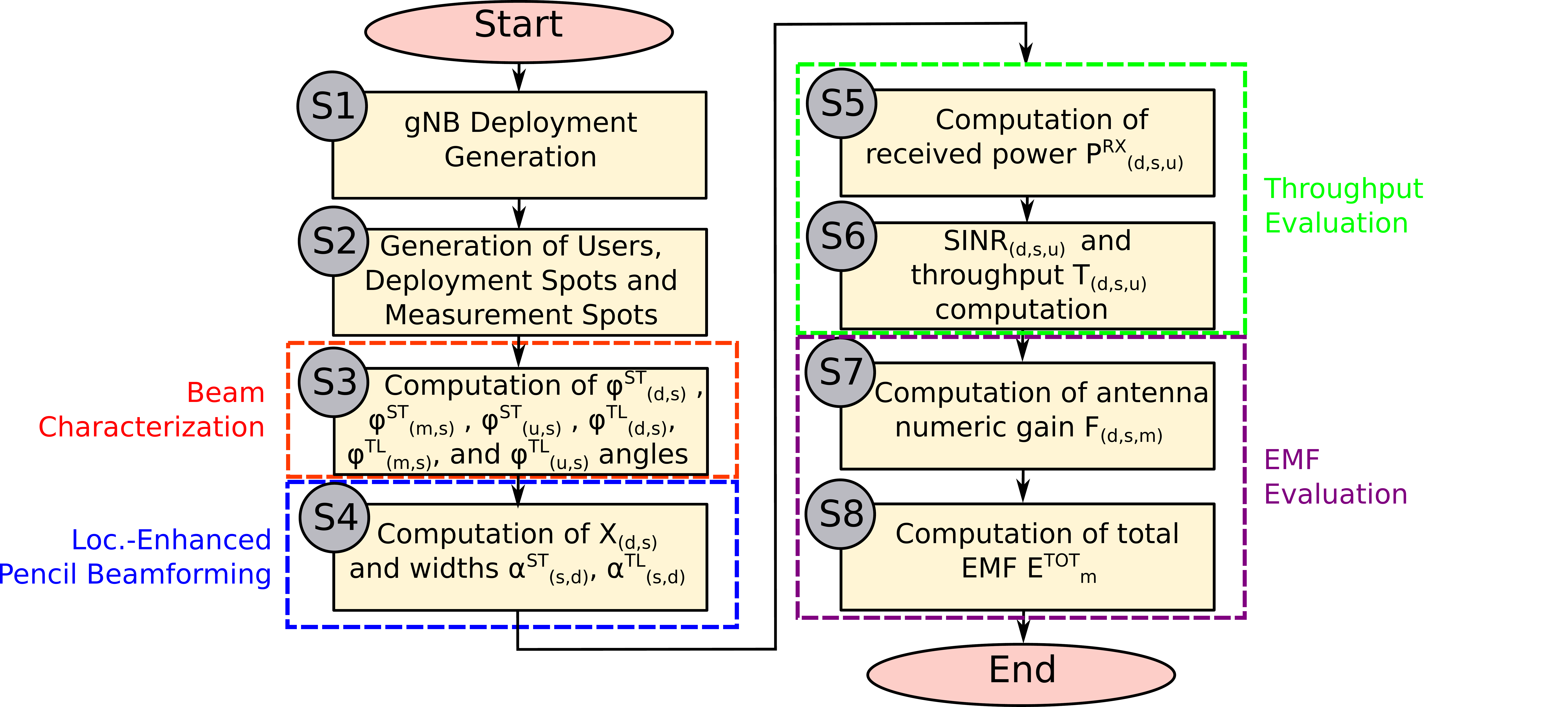}
\caption{Block diagram of the \textsc{5G-Pencil} implementation.}
\label{fig:block_diagram_mesh}
\end{figure*}

In order to better explain the further steps of the beam setting procedure, Fig.~\ref{fig:beam_aplitudes} reports a graphical sketch including one sector $s$ and one deployment spot $d$, subject to an uncertainty localization level equal to $\epsilon$. Let us consider two lines $T^{\text{E}}$ and $T^{\text{W}}$ passing from the projection of $s$ on the horizontal plane and tangent to the deployment spot shape (i.e., the circle centered in $d$ of diameter $\epsilon$). The interception point between $T^{\text{E}}$ ($T^{\text{W}}$) and the circle is denoted as $I^{\text{E}}_{(d,s)}$ ($I^{\text{W}}_{(d,s)}$). The angle centered on the projection of $s$ on the horizontal plane and spanning between $I^{\text{E}}_{(d,s)}$ and $I^{\text{W}}_{(d,s)}$ represents $\alpha^{\text{ST}}_{(d,s)}$. Let us now consider a straight line $T^{\text{C}}$ starting from the projection of $s$ on the horizontal plane and passing through $d$. The interception points between $T^{\text{C}}$ and the circular area delimiting the deployment spot are then denoted with $I^{\text{N}}_{(d,s)}$ and $I^{\text{S}}_{(d,s)}$, respectively. 
The angle centered in $s$ on the vertical plane and spanning between $I^{\text{N}}_{(d,s)}$ and $I^{\text{S}}_{(d,s)}$ denotes $\alpha^{\text{TL}}_{(d,s)}$.

Up to this point, a natural question is then: How to compute $\alpha^{\text{ST}}_{(d,s)}$ and $\alpha^{\text{TL}}_{(d,s)}$? To answer this question, we preliminary identify the triangle of edges
$I^{\text{W}}_{(d,s)}\xrightarrow{\text{H}}(s)\xrightarrow{\text{H}}I^{\text{E}}_{(d,s)}\xrightarrow{\text{H}}I^{\text{W}}_{(d,s)}$,  where the $(\cdot) \xrightarrow{\text{H}} (\cdot)$ notation denotes the two endpoints of the edge on the horizontal plane.
In a similar way, we identify on the vertical plane the triangle of edges $I^{\text{N}}_{(d,s)} \xrightarrow{\text{V}} (s)$, $I^{\text{S}}_{(d,s)} \xrightarrow{\text{V}} (s)$ and $\epsilon$. Given the triangle edges, we then apply the law of cosines (a.k.a. al-Kashi's theorem or Carnot's theorem) to compute $\alpha^{\text{ST}}_{(d,s)}$ and $\alpha^{\text{TL}}_{(d,s)}$, as detailed in lines (10-20) of Alg.~\ref{alg:pseudocode}. Clearly, the algorithm ends when all the pairs $(d,s)$ have been analyzed.

Focusing on Alg.~\ref{alg:pseudocode}, the beam width is scaled in accordance with the position of $d$ and with the uncertainty localization level $\epsilon$. Intuitively, when a deployment spot is close to the serving \acp{gNB}, the beam width will be higher than the one of the spots located at the sector edge. In a similar way, the beam width is decreased when the localization uncertainty level is reduced, thus allowing the synthesis of narrower pencil beams.

\section{\textsc{5G-Pencil} Implementation}
\label{sec:implementation}

We code \textsc{5G-Pencil} as an open-source simulator,  which is publicly available for download \cite{5gpencil}.\footnote{The simulator is currently avaialble for download from Dropbox. Upon acceptance of the paper, we will upload the simulator to relevant code repositories (e.g., GitHub).}
Fig.~\ref{fig:block_diagram_mesh} reports the block diagram of our implementation. During step S1, the simulator generates a candidate set of \acp{gNB} from a given coverage tessellation and a given number of sectors for each \ac{gNB}. This phase also includes the definition of the exclusion zone for each \ac{gNB}. During step S2, we generate the deployment spots, the real \ac{UE} locations, and the measurement spots. The number of deployment spots per sector is an input parameter, bounded to $N^\text{R}_s$, i.e., the number of radiating elements of each antenna array. The spatial positioning of deployment spots on the horizontal plane integrates multiple options, which include, e.g., an uniform positioning across the sector extent or a preferential generation of the spots in proximity of the sector (in order to mimic a hot-spot zone). For each deployment spot $d$, we randomly generate the real \ac{UE} location as a point inside the circle of diameter $\epsilon$, centered in $d$. Focusing then on the measurement spots, an uniform grid of equally spaced deterministic points is assumed over the whole covered territory (except inside the \acp{gNB} exclusion zone). During step S3, the simulator computes the angles $\phi^{\text{ST}}_{(d,s)}$, $\phi^{\text{TL}}_{(d,s)}$, $\phi^{\text{ST}}_{(m,s)}$, $\phi^{\text{TL}}_{(m,s)}$, $\phi^{\text{ST}}_{(u,s)}$, $\phi^{\text{TL}}_{(u,s)}$, for each deployment spot $d \in \mathcal{D}$, each measurement spot $m \in \mathcal{M}$, each real \ac{UE} location, and each sector $s \in \mathcal{S}$. In the following step (S4), the localization-enhanced pencil beamforming algorithm is executed. Consequently, both the variable $X_{(d,s)}$ and the beam widths $\alpha^{\text{ST}}_{(d,s)}$ and $\alpha^{\text{TL}}_{(d,s)}$ are set. During S5-S6, the throughput is evaluated, by adopting the model detailed in Sec.~\ref{subsec:thr_eval}. Finally, the \ac{EMF} of the pencil beam is computed in S7-S8, by adopting the procedure described in Sec.~\ref{subsec:emf_eval}.


\begin{table}[t]
    \caption{Settings of the Main Input Parameters.}
    \label{tab:notation}
    \scriptsize
    \centering
    \begin{tabular}{|p{0.6cm}|p{3cm}|p{3.5cm}|}
\hline
 \rowcolor{Coral}  \textbf{Symbol} & \textbf{Description} & \textbf{Value/Reference} \\
\hline
 - & \begin{minipage}{3cm}\ac{gNB} Deployment\end{minipage} & \begin{minipage}{3.5cm}Hexagonal grid\end{minipage} \\
\rowcolor{Linen} $N^{\text{gNB}}$ & \begin{minipage}{3cm}Number of \acp{gNB}\end{minipage} & 7 \\
 $N^{\text{SEC}}$ & \begin{minipage}{3cm}Number of sectors per \ac{gNB}\end{minipage} & \begin{minipage}{3.5cm}3 (with 120$^\circ$ orientation)\end{minipage} \\
\rowcolor{Linen} $L$ & \begin{minipage}{3cm}Sector side\end{minipage} & 100~m \\
 $N^{\text{R}}_s$ & \begin{minipage}{3cm}Number of radiating elements per sector\end{minipage} & 64 \cite{gnbdatastheets} \\
\rowcolor{Linen} $|\mathcal{S}|$ & \begin{minipage}{3cm}Total number of sectors\end{minipage} & \begin{minipage}{3.5cm}$N^{\text{gNB}} \cdot N^{\text{SEC}} = 21$\end{minipage}  \\
 $|\mathcal{D}|$ &  \begin{minipage}{3cm}Number of deployment spots\end{minipage} & \begin{minipage}{3.5cm}$|\mathcal{S}|\cdot N^{\text{R}}_s=448$ \end{minipage}  \\
\rowcolor{Linen} - & \begin{minipage}{3cm}Deployment spot positioning\end{minipage} & \begin{minipage}{3.5cm}Hot-spot placement with random polar coordinates\end{minipage}  \\ 
 & & \\[-0.9em]
 - & \begin{minipage}{3cm}Coverage check\end{minipage} & \begin{minipage}{3.5cm}Based on Voronoi region (with sectorization)\end{minipage}\\
\rowcolor{Linen} $|\mathcal{U}|$ &  \begin{minipage}{3cm}Number of real \ac{UE} locations\end{minipage} & \begin{minipage}{3.5cm}$N^{\text{SEC}} \cdot N^{\text{R}}_s=192$ (central \ac{gNB})\end{minipage}  \\
 - & \begin{minipage}{3cm}Real \ac{UE} location positioning\end{minipage} & \begin{minipage}{3.5cm}Random placement based on Cartesian coordinates in the circle centered in $d$ with radius $\epsilon$\end{minipage} \\
 & & \\[-0.9em]
\rowcolor{Linen}  $|\mathcal{M}|$ &  \begin{minipage}{3cm}Number of measurement spots\end{minipage} & \begin{minipage}{3.5cm} 25572 (central \ac{gNB} with $1~\text{m} \times 1~\text{m}$ resolution)  \end{minipage}  \\
 $R^{\text{MIN}}_s$ & \begin{minipage}{3cm}Exclusion Zone Radius\end{minipage} & 10~m \cite{Tornevik} \\
\rowcolor{Linen} $P^{\text{TX}}_{s}$ & \begin{minipage}{3cm}Max. TX power per antenna array\end{minipage} & 200~W \cite{Tornevik}\\ 
$P^{\text{MAX}}_s$ & \begin{minipage}{3cm}Power of one antenna element\end{minipage} & \begin{minipage}{3cm}$P^{\text{TX}}_{s}/N^{\text{R}}_s=3.125$~W (Uniform power splitting among the antenna elements)\end{minipage}  \\
\rowcolor{Linen} $G^{\text{MAX}}$  & \begin{minipage}{3cm}Maximum antenna gain\end{minipage} & 15~dBi \cite{gnbdatastheets}\\
 $A^{\text{AZ}}_{\text{MIN}}$ & \begin{minipage}{3cm}Front-to-back ratio\end{minipage} & 25~dB \cite{yu2016load} \\
\rowcolor{Linen} $A^{\text{EL}}_{\text{MIN}}$ & \begin{minipage}{3cm}Side lobe limit\end{minipage} & 20~dB \cite{yu2016load} \\
 $B_s$ & \begin{minipage}{3cm}Sector bandwidth\end{minipage} & 80~MHz \cite{italianauction} \\
\rowcolor{Linen} $F_s$ &\begin{minipage}{3cm}Sector operating frequency\end{minipage}& 3.7~GHz \cite{italianauction} \\
 $H_s$ &\begin{minipage}{3cm}Sector height above ground\end{minipage}& 15~[m]\\
\rowcolor{Linen}$H_d$ &  &  \\
\rowcolor{Linen} $H_m$ &  &  \\
\rowcolor{Linen} $H_u$ & \multirow{-3}{*}{\begin{minipage}{3cm}Deployment spot/measurement spot/\ac{UE} height above ground\end{minipage}} & \multirow{-3}{*}{1.5~[m]} \\
  $L^{\text{PL}}_{(s,u)}$ & \begin{minipage}{3.5cm}3D path loss between $s$ and $u$\end{minipage} & 
\begin{minipage}{3cm}3GPP UMi-Street Canyon LOS/NLOS models \cite{pathlossmodel}\end{minipage} \\

\rowcolor{Linen} $G^{\text{TX}}_{s}$ & \begin{minipage}{3cm}TX gain per antenna element\end{minipage} & 3~dBi  \cite{awada2017simplified}\\
 $G^{\text{BF}}_{s}$ & \begin{minipage}{3cm}Maximum beamforming gain\end{minipage} & \begin{minipage}{3cm}$10 \cdot \log_{10}(N^{\text{R}}_s)$ \cite{awada2017simplified}\end{minipage} \\
\rowcolor{Linen} $N_s$ & \begin{minipage}{3cm}Noise term\end{minipage} & \begin{minipage}{3cm}Noise power from \cite{noisesetting} with 5~dB noise figure and $B_s$ bandwidth\end{minipage}\\
 $\alpha^{\text{ST}}_{\text{MIN}}$ & \begin{minipage}{3cm}Minimum width angle (steering)\end{minipage} & 3$^\circ$ \\
\rowcolor{Linen} $\alpha^{\text{TL}}_{\text{MIN}}$ & \begin{minipage}{3cm}Minimum width angle (tilting)\end{minipage} & 3$^\circ$ \\
 $\epsilon$ & \begin{minipage}{3cm}Uncertainty localization level\end{minipage} & \begin{minipage}{3.5cm}$\{20,16,8,4,2\}$~m (Range covering PSL 1-6 \cite{3GPP:TS:22.261:V18.1.0}) \end{minipage}\\  
\hline
\end{tabular}
\end{table}

\section{Scenario Description}
\label{sec:scenario}

\begin{figure}[t]
\centering
\includegraphics[width=7cm]{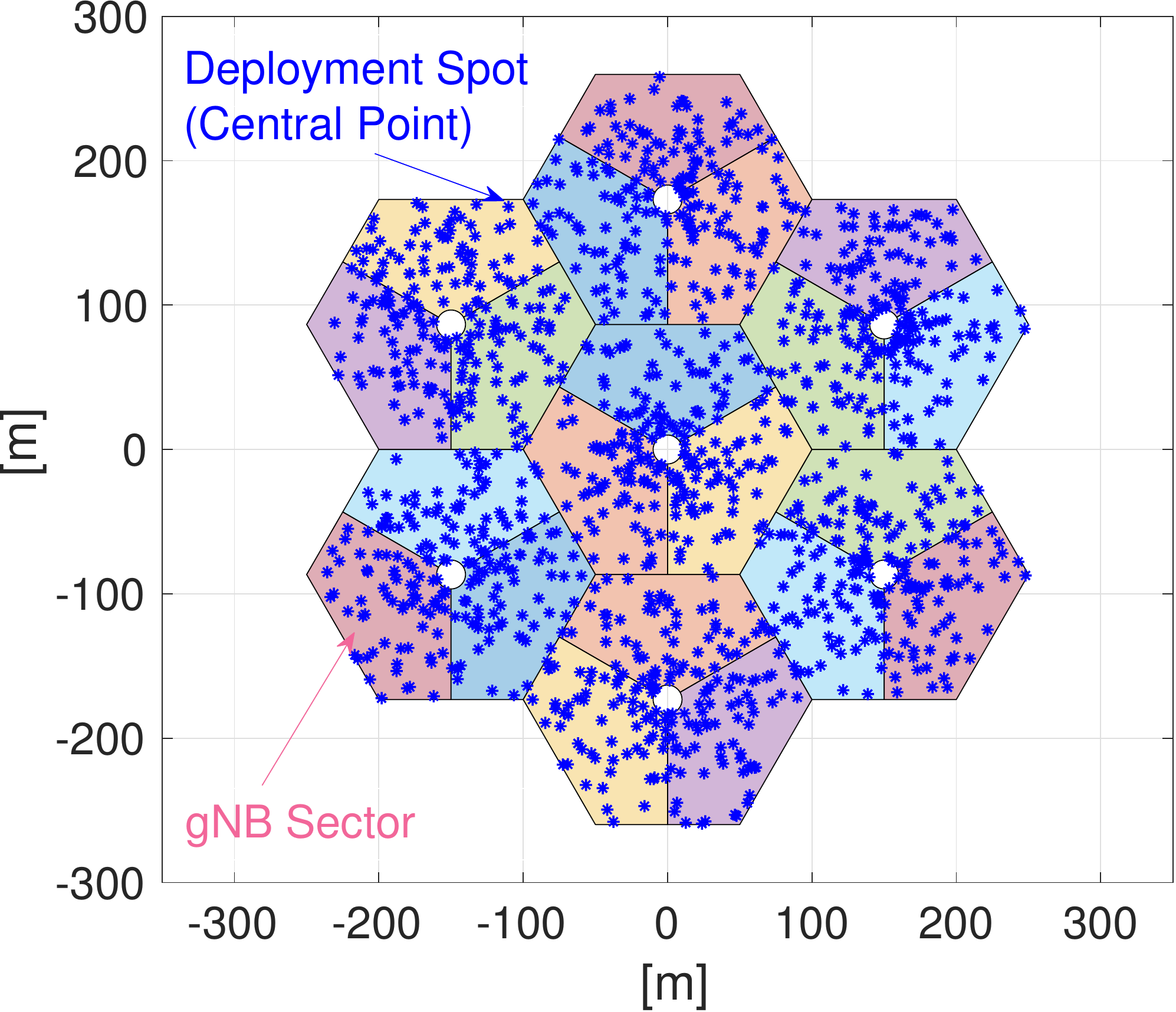}
\caption{Example of \acp{gNB} and deployment spots positioning in \textsc{5GPencil}.}
\label{fig:example_deployment}
\vspace{-4mm}
\end{figure}

We consider a simple, yet meaningful scenario, to evaluate the impact of pencil beamforming that is achieved by running the \textsc{5G-Pencil} framework. To this aim, Tab.~\ref{tab:notation} reports the settings for the main input parameters. More specifically, we consider a regular cellular deployment, where a set of $N^{\text{gNB}}=7$ \acp{gNB} are placed on an hexagonal grid. We then assume that the coverage area of each \ac{gNB} is an hexagon of side $L=100$~[m], thus matching an urban/dense urban 5G deployment. Moreover, each \ac{gNB} is equipped with $N^{\text{SEC}}=3$ non-overlapping sectors, with a circular exclusion zone of radius $R^{\text{MIN}}_s=10$~[m] around the \ac{gNB} location (set in accordance with \cite{Tornevik}). The total number of sectors $|\mathcal{S}|$ is then equal to $N^{\text{gNB}} \cdot N^{\text{SEC}} = 21$. Each sector is equipped with $N^{\text{R}}_s=64$ radiating elements, as reported by relevant datasheets of 5G equipment \cite{gnbdatastheets}. Focusing then on the deployment spots generation, we assume that each sector simultaneously serves the maximum number of spots, which corresponds to $N^{\text{R}}_s=64$. Consequently, the total number of deployment spots over all the sectors is equal to $|\mathcal{S}|\cdot N^{\text{R}}_s=448$. The set of deployment spots in each sector is generated by randomly picking polar coordinates over the sector extent, which then results in a set of spots preferentially generated in the surroundings of the sector center. Finally, Fig.~\ref{fig:example_deployment} a graphical sketch of the considered set of sectors, as well as one exemplary generation of deployment spots.

In the following, we move our attention on the selection of the zones in which the \ac{EMF} and the throughput are evaluated. Focusing on the former, we deploy an uniform grid of squared measurement spots with resolution of 1~[m] $\times$ 1~[m], covering the area of the central \ac{gNB} (outside the \ac{gNB} exclusion zone). In this way, each measurement spot receives exposure from both the serving sector as well as the neighboring ones (from other \acp{gNB}), while limiting the border effects that may emerge in the outer sectors. Consequently, the \ac{EMF} is evaluated over more than 25000 measurement spots. In a similar way, we restrict the throughput evaluation only for the area covered by the central \ac{gNB}. In this area, we randomly generate each \ac{UE} location within the circle centered in $d$ (corresponding to the estimated \ac{UE} location), with radius $\epsilon$. Consequently, the throughput is evaluated over more than 190 \ac{UE} locations.

We then  focus on the parameters that are needed by the \ac{EMF} and throughput models. We refer the reader to Tab.~\ref{tab:notation} for the detailed explanation of each parameter setting, while here we discuss the salient features. In brief, most of parameters are taken from the literature, product datasheets and/or real deployment options. Focusing on the bandwidth and frequency, we consider the deployment of \acp{gNB} operating in the mid-band (i.e., 3.7~[GHz] of operating frequency), with 80~[MHz] of available bandwidth for each sector.\footnote{5G also includes sub-GHz and mm-Waves frequencies, which are however intentionally not treated in this work, due to the following reasons. First, it is expected that \acp{gNB} operating on sub-GHz frequencies will be mainly used for coverage. Therefore, the benefits of pencil beamforming will be limited in this case. Second, the installation of \acp{gNB} operating on mm-Waves is still at an early-stage in many contries, and mostly confined to specific scenarios (e.g., very dense areas). Consequently, in this work we focus on \acp{gNB} operating on mid-band frequencies, which are the currently adopted option for realizing 5G in many countries in the world (including Italy). The investigation of pencil beamforming with mm-Waves \acp{gNB} is left for future work.} In addition, we assume that the total power consumption of the antenna array is uniformly split across the radiating elements. Consequently, the power of one antenna element $P^{\text{MAX}}_s$  is set equal to 3.125~[W]. As an additional comment, we consider the 3GPP UMi-Street Canyon propagation model \cite{pathlossmodel} under both \ac{LOS} and \ac{NLOS} conditions. To this aim, each \ac{gNB} sector is placed at height of 15~[m] above ground level (corresponding to a pole-mounted and/or roof-top installation), while the \ac{EMF} and the throughput are evaluated at an height of 1.5~[m] above ground level.

Focusing then on the technology constraints to synthesize the beams, we have to select the minimum beam widths $\alpha^{\text{ST}}_{\text{MIN}}$ and $\alpha^{\text{TL}}_{\text{MIN}}$. Although the exact settings of such parameters is strongly influenced by the adopted beamforming architecture (see e.g., \cite{beamformers}), in this work we consider $\alpha^{\text{ST}}_{\text{MIN}}=\alpha^{\text{TL}}_{\text{MIN}}=3^\circ$, due to the following reasons. First, such setting is inline with other relevant works targeting beam management in 5G networks (see e.g., \cite{fazliu2020mmwave}). Second, it is expected that 5G \ac{gNB} adopting pencil beamforming will be able to synthesize very narrow traffic beams (i.e., in the order of few degrees).

\newcolumntype{P}[1]{>{\centering\arraybackslash}m{#1}}	

\begin{table}[t]
    \scriptsize
 \begin{center}
    \caption{Uncertainty Localization Level $\epsilon$ for Different Positioning Service Levels (PSLs) \cite{3GPP:TS:22.261:V18.1.0}.}
    \label{tab:table1}
{  \begin{tabular}{|P{1cm}| P{2cm}|P{3cm}|P{1cm}|} 
\hline
    \rowcolor{Coral} \textbf{Level index} & \textbf{Coverage} & \textbf{Deployment} & $\epsilon$\\
     \hline
1		&	 Indoor/Outdoor & Rural/Urban &	20~m\\
\rowcolor{Linen} 2		    & 	Outdoor &
Rural/Urban/Dense urban &	6~m\\
\{3,4,5,6\}  & 	Indoor/Outdoor  & Rural/Urban/Dense urban &	4~m\\
\hline
    \end{tabular} }
  \end{center}\vspace{-0.4cm}
\end{table}

Finally, we take into account the setting for one of the most impacting parameters: the localization uncertainty level $\epsilon$. To this aim, we rely on \ac{PSL} 1-6 defined in \cite{3GPP:TS:22.261:V18.1.0}, which correspond to the accuracy requirements that 5G networks should fulfill according to the 3GPP definition of location services.\footnote{We intentionally neglect level 7 from our analysis, since the accuracy for such level  is intended for relative positioning instead of absolute positioning.} Intuitively, each \ac{PSL} is characterized by given values of horizontal/vertical accuracy and absolute/vertical position, by considering other timing \acp{KPI}, as well as by taking into account the different operating environments and network coverage. Tab.~\ref{tab:table1} summarizes the aforementioned \acp{PSL} from \cite{3GPP:TS:22.261:V18.1.0}, by reporting the values of $\epsilon$ (corresponding to the horizontal accuracy requirement). In our work, we consider a range of $\epsilon$ values that covers the range of values of Tab.~\ref{tab:table1}. Consequently, we selectively set $\epsilon=\{20,16,8,4,2\}$~m.

\section{Results}
\label{sec:results}

\textbf{Reference Solutions.} In order to position our approach, we consider the following reference approaches: \textit{i}) beamforming with fixed widths, and \textit{ii}) no beamforming. Focusing on \textit{i}), we run S1-S8 of \textsc{5G-Pencil}, by replacing S4 with a fixed width assignment. In particular, the beam widths are set equal to fixed angles $\alpha^{\text{ST}}_{\text{FIXED}}$, $\alpha^{\text{TL}}_{\text{FIXED}}$, which are retrieved from product data-sheets \cite{gnbdatastheets} and research works \cite{ali2019system}. In this way, we assume to apply a ``soft'' beamforming, where localization is solely exploited to tune the beam direction, without tuning the beam widths in accordance with $\epsilon$. Focusing on EMF and throughput evaluation, we employ the same models already introduced in Sec.~\ref{subsec:emf_eval} and Sec.~\ref{subsec:thr_eval} (with $\alpha^{\text{ST}}_{(d,s)}=\alpha^{\text{ST}}_{\text{FIXED}}$ and $\alpha^{\text{TL}}_{(d,s)}=\alpha^{\text{TL}}_{\text{FIXED}}$), in order to perform a fair comparison.

Regarding instead the case without beamforming, we introduce such term of comparison for \ac{EMF} evaluation. More in depth, we assume that each \ac{gNB} is realized with an omnidirectional antenna, always radiating at the maximum power in all directions. As a consequence, both sectorization and beamforming are not employed in this case. More formally, we adopt a simplified version of the \ac{ITU} point-source model to compute the \ac{PD} that is received over each measurement spot $m \in \mathcal{M}$:
\begin{equation}
S^{\text{TOT}}_m=\sum_{g} \frac{P^{\text{MAX}}_g \cdot G^{\text{MAX}}}{4 \pi \cdot \delta^2_{(g,m)}}
\end{equation}
where $g$ is the considered \ac{gNB} (belonging to the set of \acp{gNB} $\mathcal{G}$), $\delta_{(g,m)}$ is the 3D distance between \ac{gNB} $b$ and measurement spot $m$ and $P^{\text{MAX}}_g$ is the maximum radiated power by a 5G \ac{gNB} (set to 200~[W] in accordance with \cite{Tornevik}). Finally, the EMF strength is computed by applying Eq.~(\ref{eq:tot_e}). As a side comment, this setting represents a very conservative case, which may result in a potential large exposure over the covered area.

\begin{figure}[t]
\centering
\includegraphics[width=8cm]{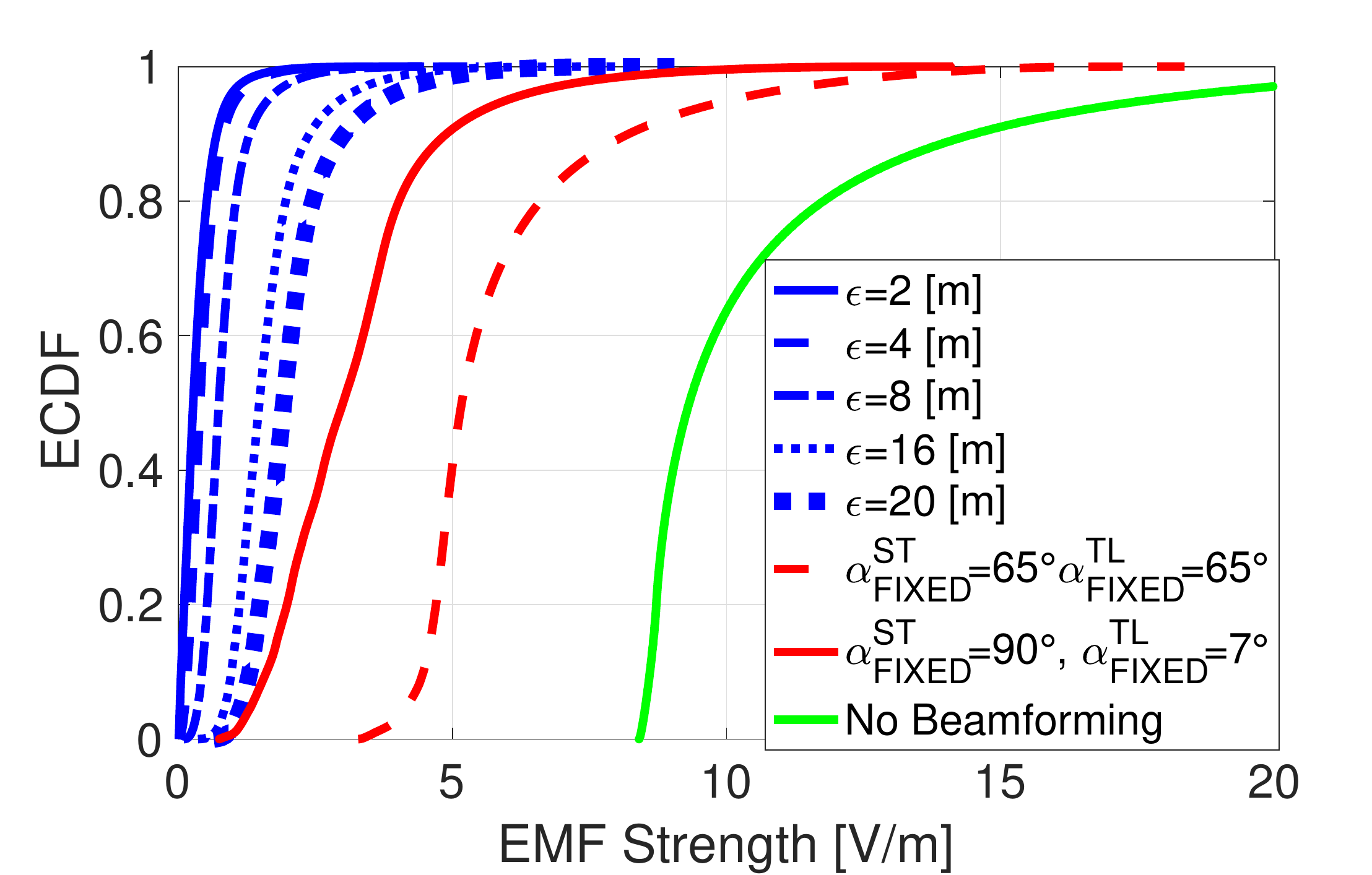}
\caption{ECDF of the average \ac{EMF} in each measurement spot by considering: \textit{i}) pencil beamforming (for different values of localization uncertainty level $\epsilon$), \textit{ii}) beamforming with fixed widths and \textit{iii}) no beamforming.}
\label{fig:ECDF_Adapt}
\end{figure}

\begin{figure}[t]
\centering
\includegraphics[width=8cm]{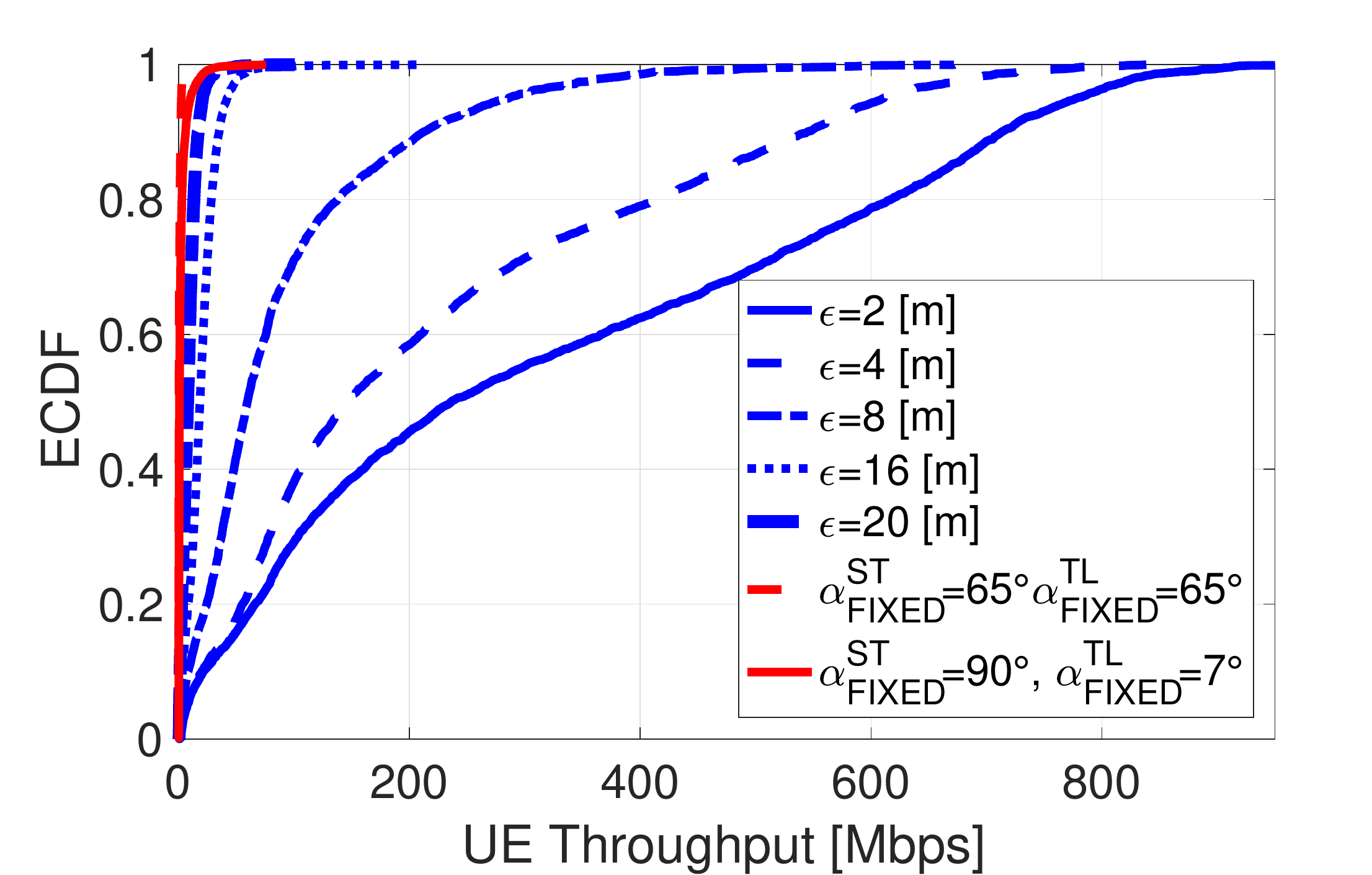}
\caption{ECDF of the users throughput by considering: \textit{i}) pencil beamforming (for different values of localization uncertainty level $\epsilon$), \textit{ii}) beamforming with fixed widths.}
\label{fig:ECDF_tp_adapt}
\end{figure}

\textbf{EMF and Throughput Comparison.}
We initially evaluate the \ac{EMF} exposure over each measurement point $m$ of the central \ac{gNB}. In this way, in fact, the total \ac{EMF} results from the exposure of the central \ac{gNB} plus the one from the six neighboring ones. In addition, we consider 20 independent runs for generating the coordinates of deployment spots in each sector $s$. We then run \textsc{5G-Pencil} to compute $E^{\text{TOT}}_{m}$ for each deployment spot generation, and then we compute the average of \ac{EMF} over the 20 runs for each $m$. At the same time, we collect the throughput value for each deployment spot $d$ that is served by the sectors of the central \ac{gNB}. Unless otherwise specified, we assume \ac{NLOS} conditions, and interference generated by the same sector (intra-sector term in Eq.~\ref{eq:sinr}) as well by the neighboring ones (inter-sector term in Eq.~\ref{eq:sinr}).   

\begin{figure}[t]
\centering
\subfigure[EMF strength, $\epsilon=20$~m]
{
	\includegraphics[width=4cm]{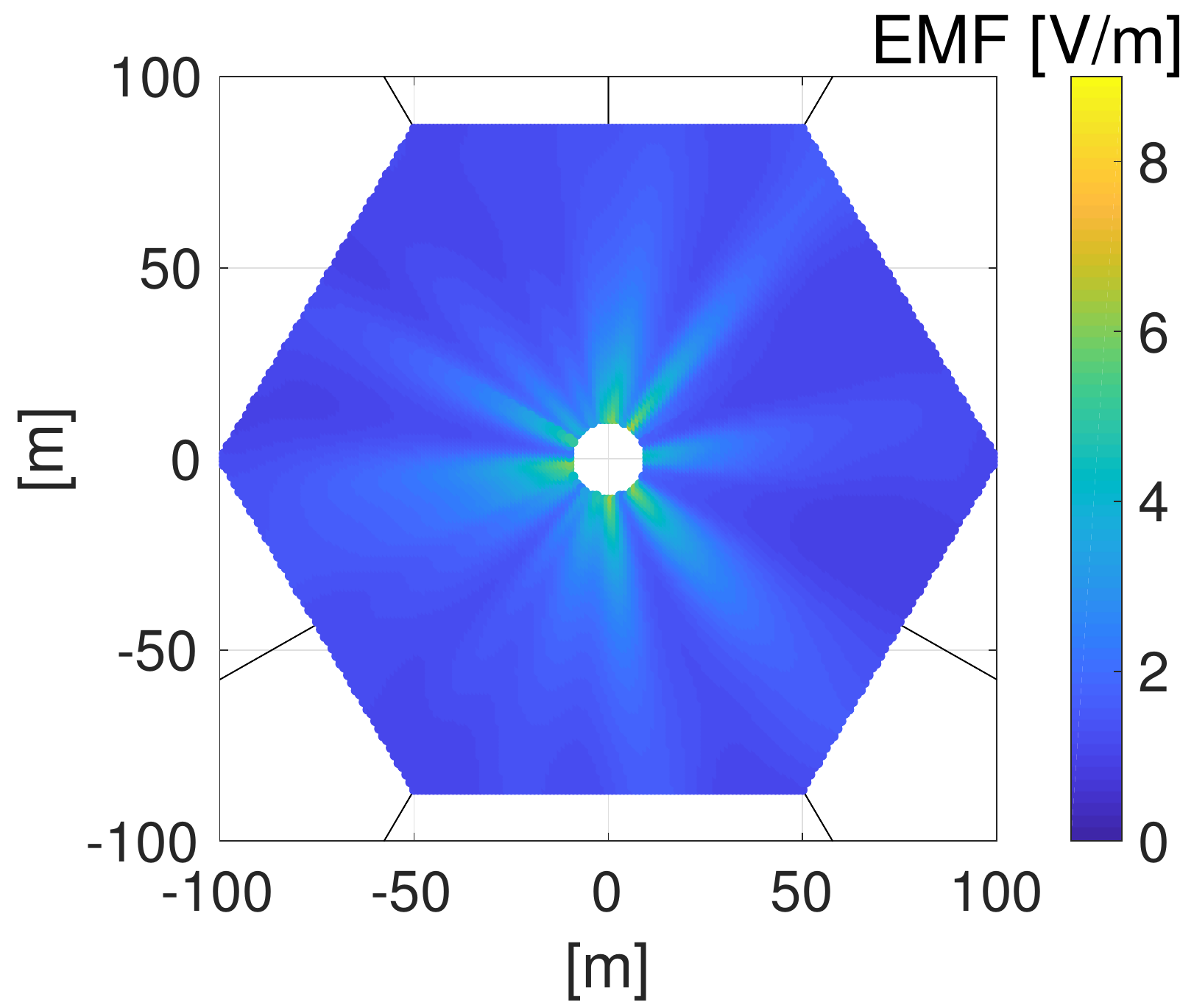}
	\label{fig:Adapt_emf_20}
}
\subfigure[EMF strength, $\epsilon=2$~m]
{
	\includegraphics[width=4cm]{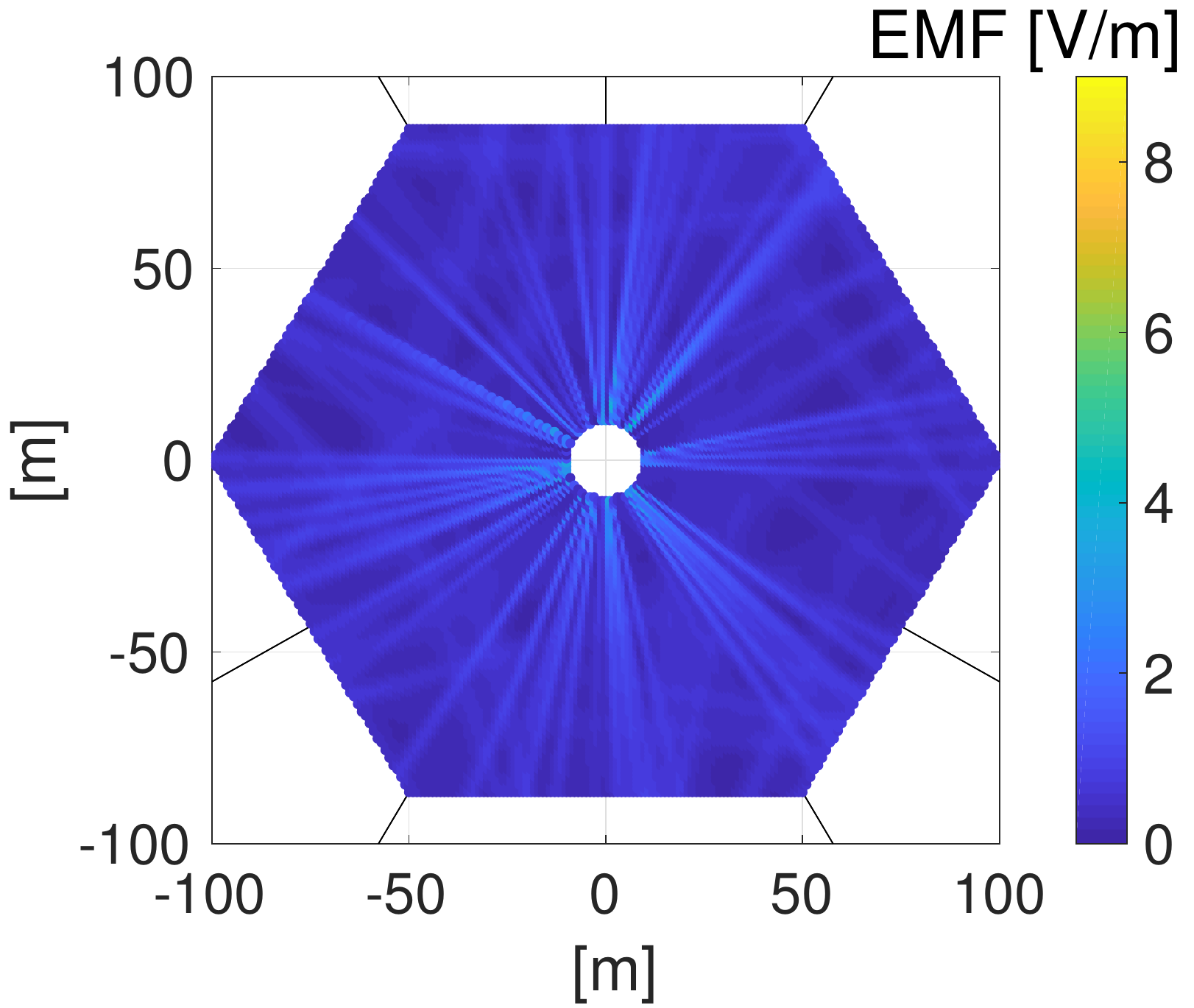}
	\label{fig:Adapt_emf_2}
}

\subfigure[Number of beams, $\epsilon=20$~m]
{
	\includegraphics[width=4cm]{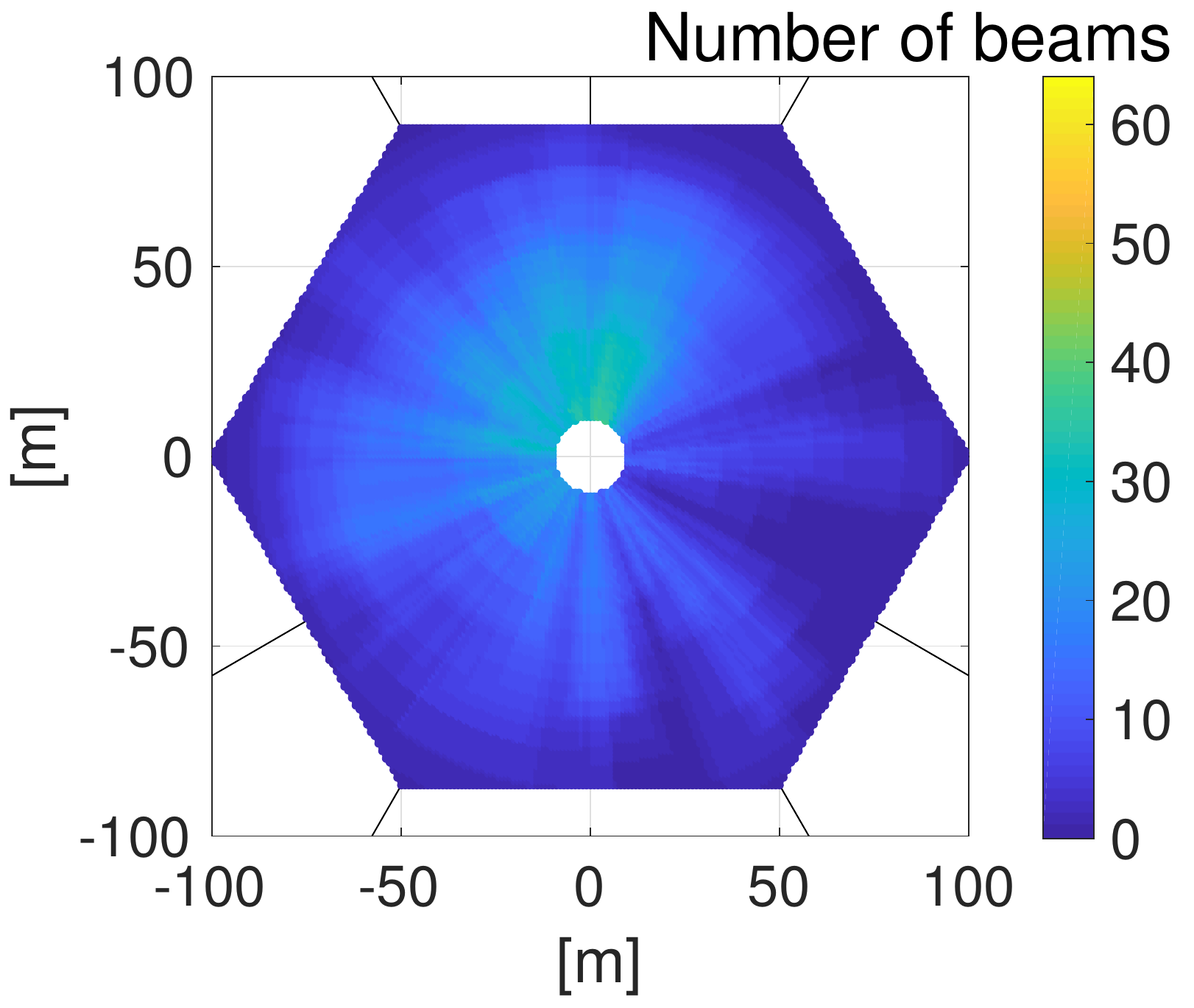}
	\label{fig:Adapt_beams_20}
}
\subfigure[Number of beams, $\epsilon=2$~m]
{
	\includegraphics[width=4cm]{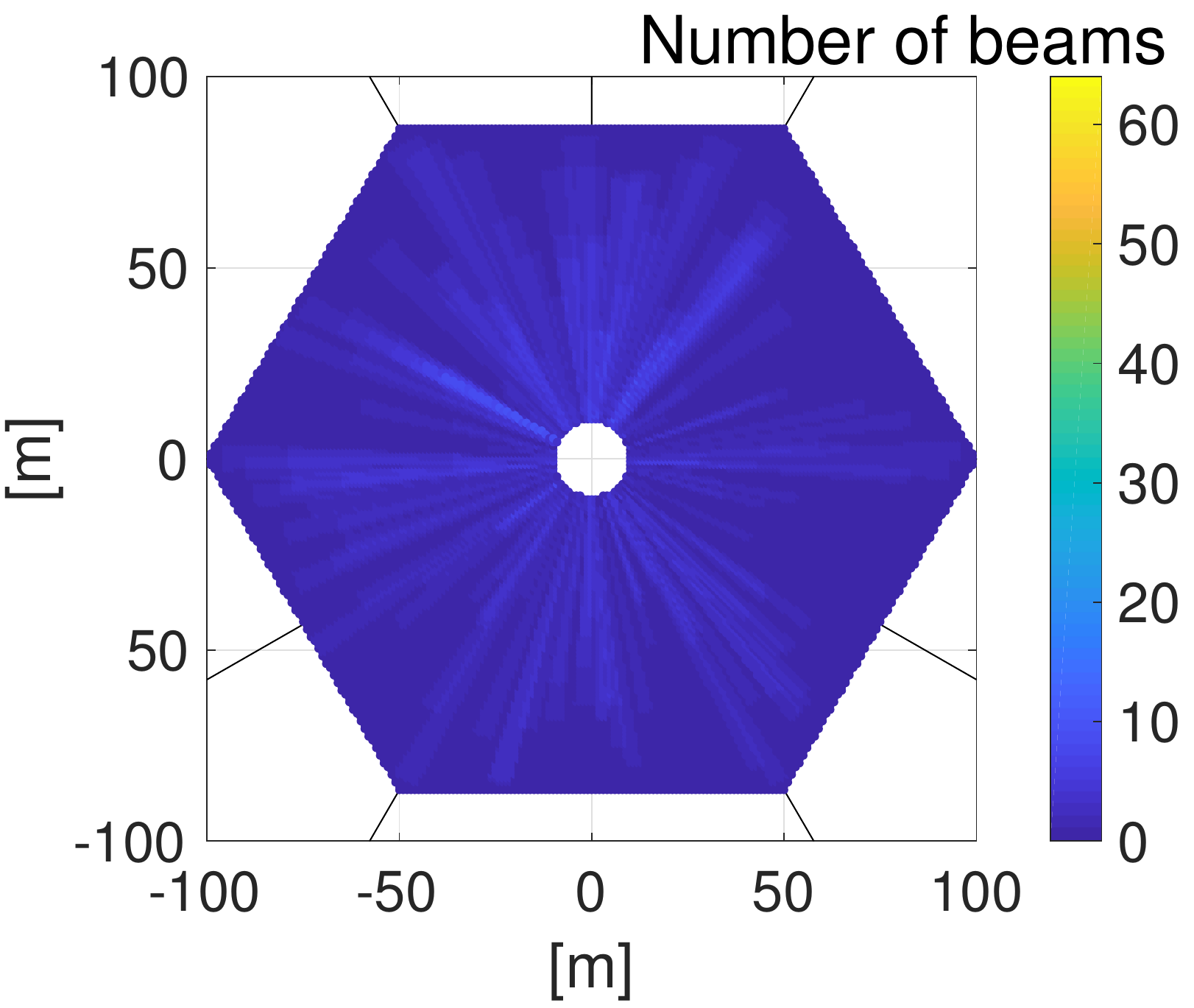}
	\label{fig:Adapt_beams_2}
}

\caption{EMF strength and number of overlapping beams for $\epsilon=\{20,2\}$~[m].}
\label{fig:Adapt_emf_beams}
\end{figure}

Fig.~\ref{fig:ECDF_Adapt} reports the \ac{ECDF} from: pencil beamforming (with different $\epsilon$ values), beamforming with fixed widths, and no beamforming cases. Several considerations hold by analyzing in detail the figure. First, the pencil beamforming functionality triggers a strong decrease of \ac{EMF} exposure compared to the reference solutions, thus contradicting the widespread belief of the population. In addition, the exposure tends to be further reduced as $\epsilon$ is decreased, due to the fact that narrower beam widths are synthesized. Eventually, the average exposure is overall pretty limited with pencil beamforming, with a maximum \ac{EMF} almost equal to 5~[V/m], i.e., a value clearly lower than the limits defined by international regulations (e.g. \ac{ICNIRP} ones \cite{international2020guidelines}).

We now evaluate the throughput in terms of \ac{ECDF} in Fig.~\ref{fig:ECDF_tp_adapt}. In this case, the throughput is evaluated under the most conservative settings, i.e., \ac{NLOS} conditions and interference including the intra-sector term. Two considerations hold by analyzing the figure. First, pencil beamforming allows achieving consistently higher throughput values compared to beamforming with fixed widths. Obviously, this improvement derives from the reduction of interference terms in Eq.~(\ref{eq:sinr}). In addition, the decrease of the localization uncertainty level $\epsilon$ results in a prompt increase of the throughput, with values even larger than 100~[Mbps] for more than 80\% of the deployment spot (with $\epsilon$=2~[m]). Therefore, we can conclude that pencil beamforming is beneficial not only in terms of \ac{EMF}, but also for the throughput levels.

\textbf{Spatial Exposure Analysis.} We then move our attention to the characterization of pencil beamforming exposure over the territory. Fig.~\ref{fig:Adapt_emf_beams} reports a high-level quantitative analysis over one run, by showing the \ac{EMF} strength and the number of overlapping beams, $\epsilon=\{20,2\}$.\footnote{To compute the overlapping beam metric, we assume a beam cone within the 3~[dB] zone, ending at point $I^{N}_{(d,s)}$.} 
For this specific test, the same positioning of deployment spot is used $\epsilon=20$~[m] and $\epsilon=2$~[m]. Interestingly, we can note that, as the localization uncertainty level is improved, both the \ac{EMF} exposure and the number of overalapping beams are reduced. This reduction is achieved not only for the measurement spots in proximity to the central \ac{gNB}, but also over the whole sector  extent. Therefore, when $\epsilon$ is decreased, the exposure over the territory tends to be reduced and in general uniformly distributed.

\begin{figure}[t]
\centering
\includegraphics[width=8cm]{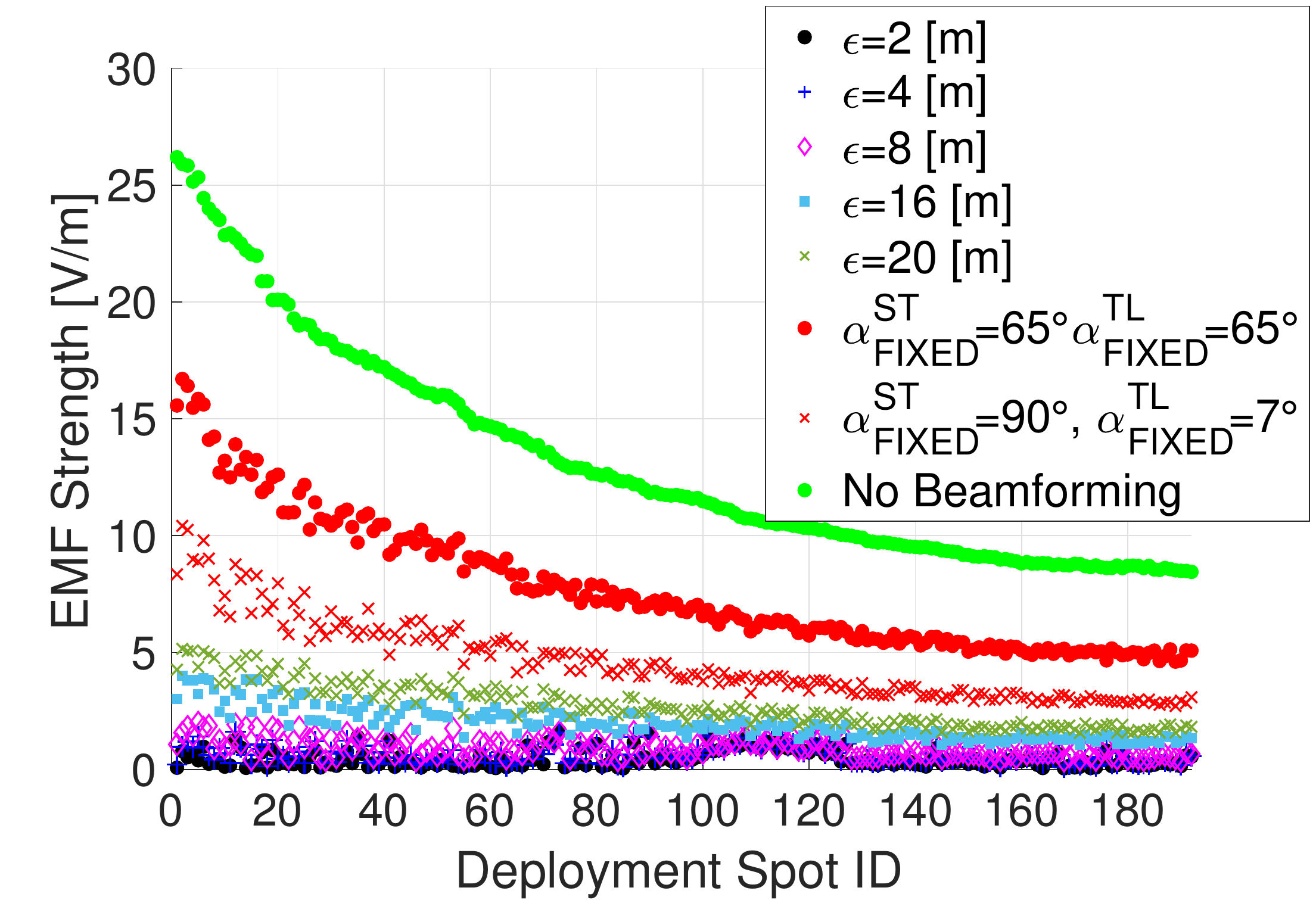}
\caption{Average \ac{EMF} in each deployment spot vs. deployment spot ID. The spots are ordered in decreasing distance w.r.t. serving sector.}
\label{fig:dp_eval}
\end{figure}

In the following step, we focus on the spatial evaluation of exposure over the deployment spots, which we remind are the zones of the territory which include \ac{UE}. In more detail, we consider one generation of deployment spots. For each spot $d \in \mathcal{D}$, we compute the \ac{EMF} as a linear average of the measurement spots falling in the circle centered in $d$ of radius $\epsilon$. We also introduce here the reference approaches, by computing the \ac{EMF} for each $d$ within a circle of radius $\epsilon=2$~[m]. Fig.~\ref{fig:dp_eval} reports the \ac{EMF} strength vs. the deployment spots ID. Each ID is uniquely assigned by considering a sorting of the spots based on decreasing distance from the serving sector. By analyzing the figure, we can note that pencil beamforming allows reducing the exposure over the deployment spots. Clearly, such reduction is higher for those spots in proximity to the radiating \ac{gNB}, i.e., those ones with lower IDs. In addition, the decrease of $\epsilon$ further reduces the exposure for all the spots w.r.t. the other reference solutions.


%

\begin{figure}[t]
\includegraphics[width=8cm]{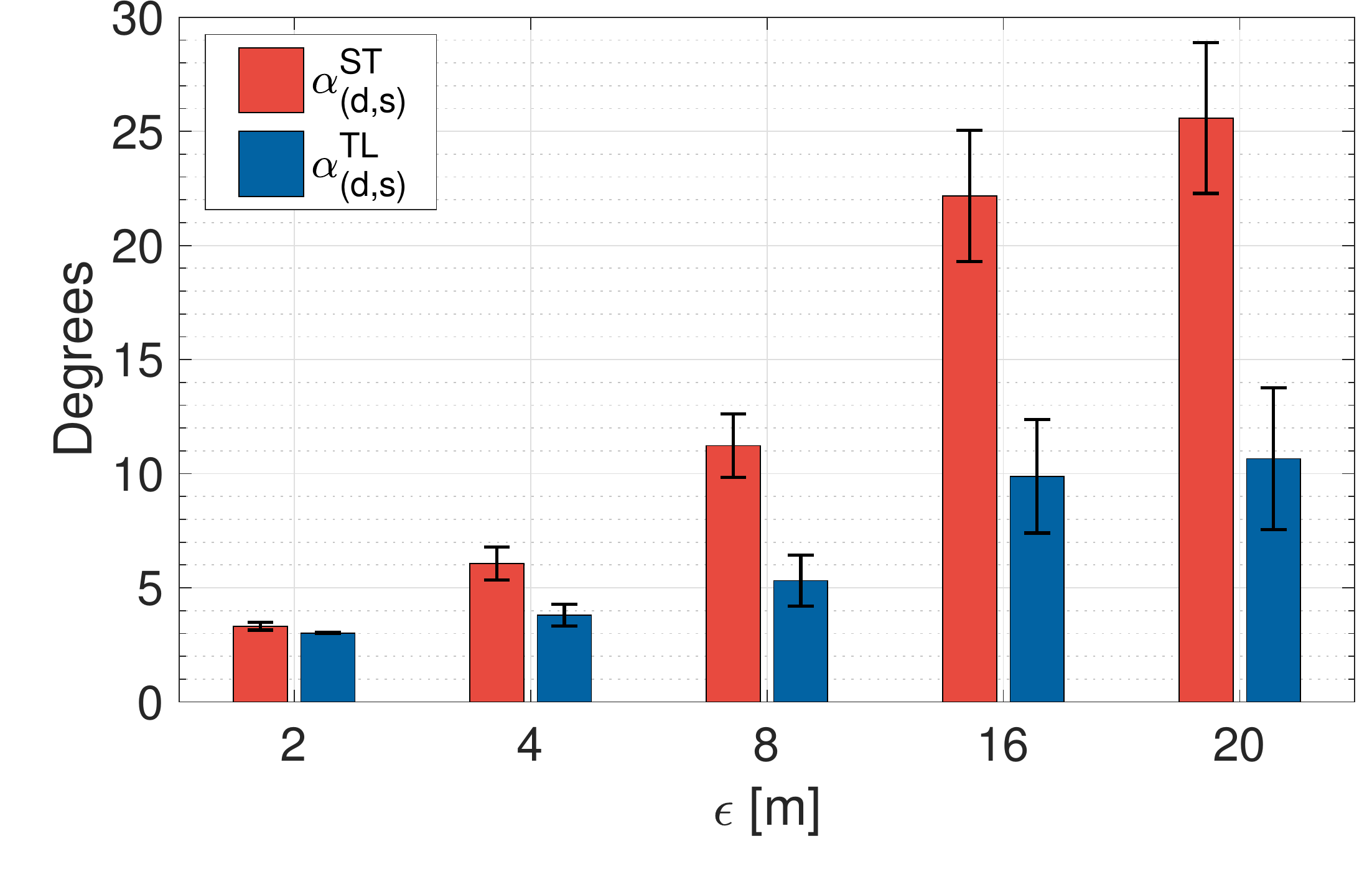}
\caption{Average values and 95\% confidence intervals for $\alpha^{\text{ST}}_{(d,s)}$ and $\alpha^{\text{TL}}_{(d,s)}$ vs. variation of localization uncertainty level $\epsilon$.}
\label{fig:angles_comp}
\end{figure}

\textbf{Beam width feasibility.}
A natural question emerges at this point: Are the angles enforced by pencil beamforming overall feasible? To answer this question, Fig.~\ref{fig:angles_comp} reports the average values and 95\% confidence intervals of $\alpha^{\text{ST}}_{(d,s)}$ and $\alpha^{\text{TL}}_{(d,s)}$ (computed over the 20 runs, for all sectors) vs. the variation of $\epsilon$. Interestingly, we can note that the beam widths tend to notably decrease as $\epsilon$ is reduced. In addition, the beam steering is always higher than the beam tilting (as expected). Clearly, the imposed angles are always higher than the minimum ones ($\alpha^{\text{ST}}_{\text{MIN}}=\alpha^{\text{TL}}_{(\text{MIN}}=3^\circ$)

\begin{figure}[t]
\centering
\subfigure[EMF Evaluation]
{
	\includegraphics[width=4.1cm]{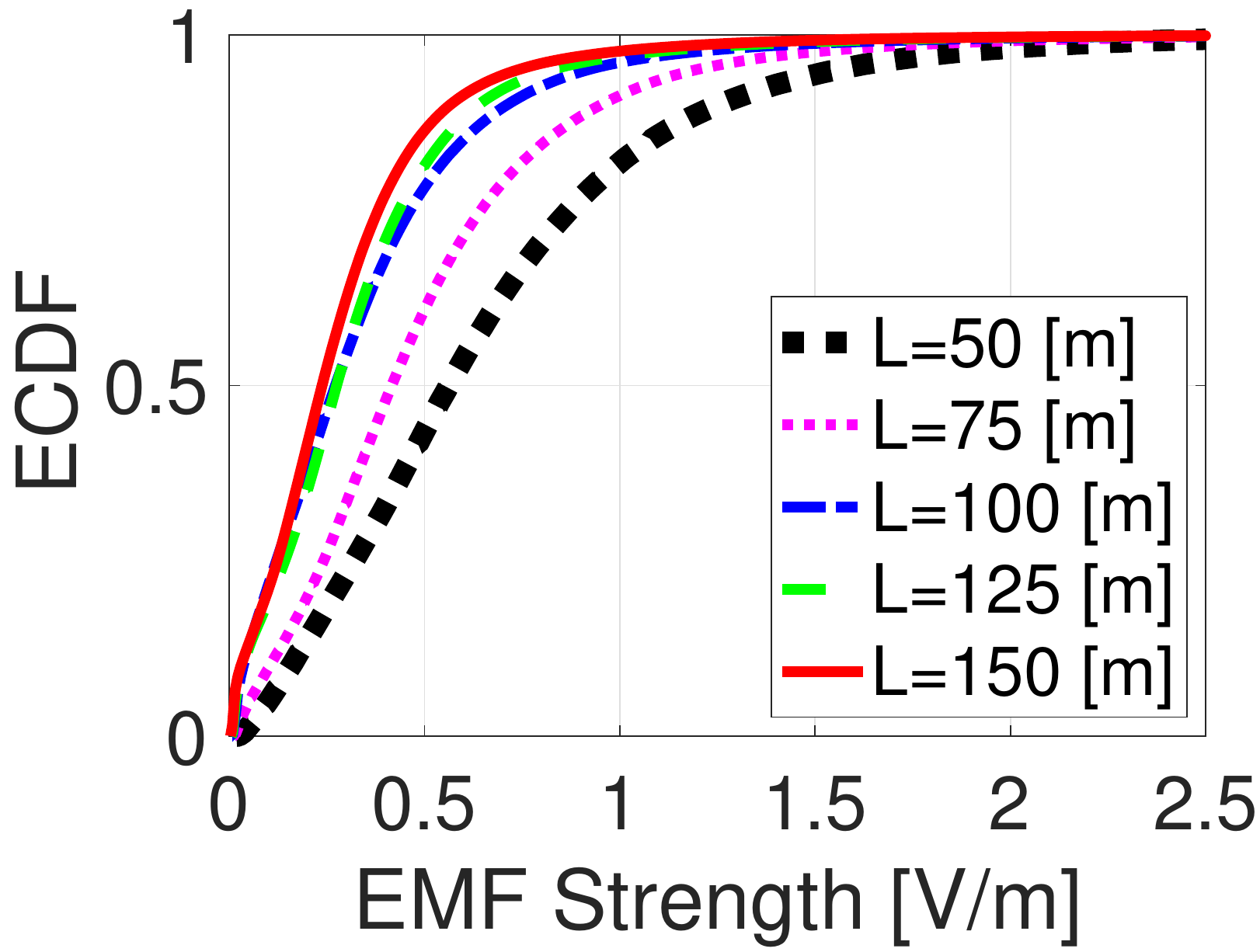}
	\label{fig:emf_eval_var_l_constant_dp}
}
\subfigure[Throughput Evaluation]
{
	\includegraphics[width=4.1cm]{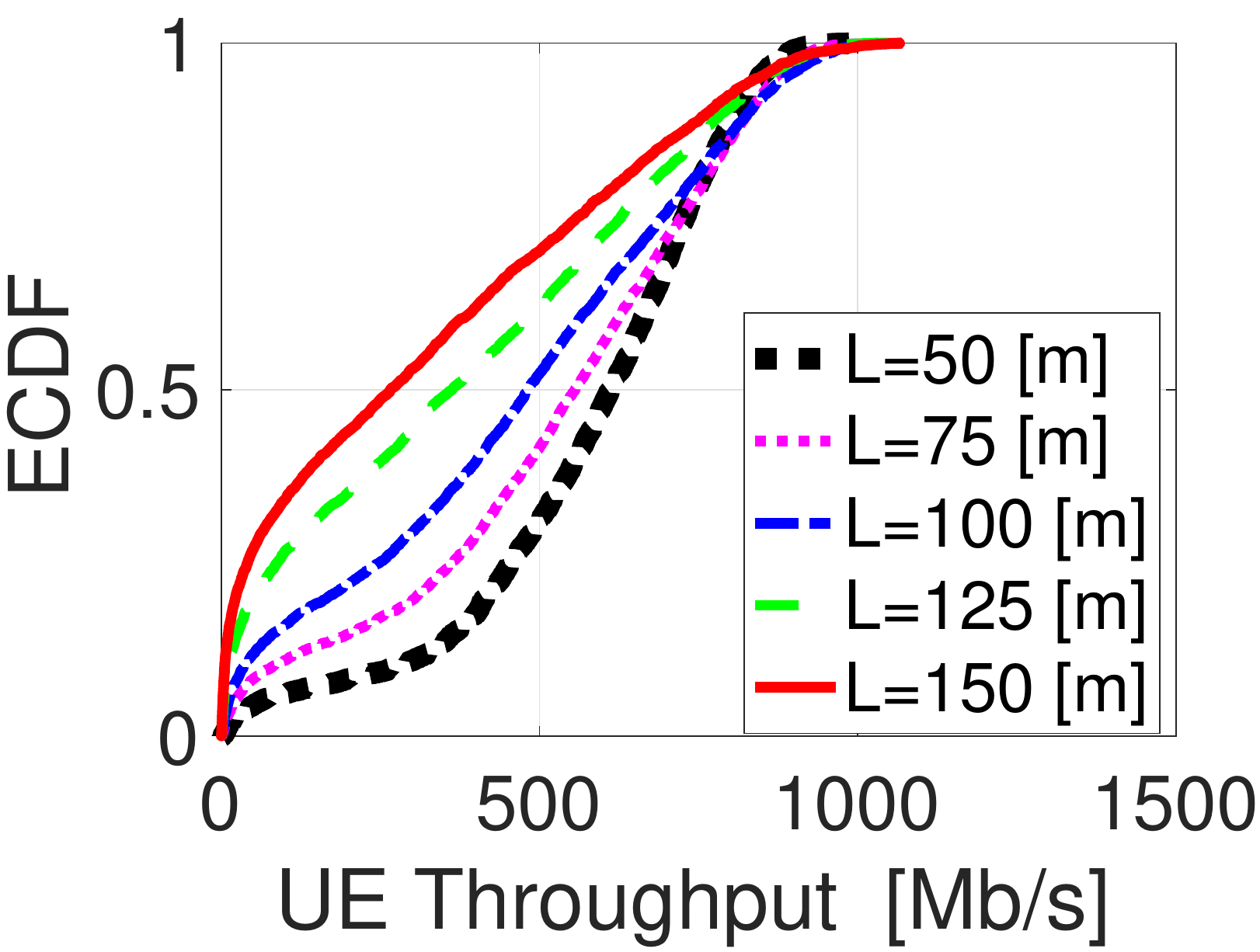}
	\label{fig:throughput_eva_var_l_constant_dp}
}
\caption{EMF and throughput evaluation vs. variation of sector size $L$ (\ac{UE} density decreasing with $L$).}
\label{fig:var_param_dp_constant}
\end{figure}

\begin{figure}[t]
\centering
\subfigure[EMF Evaluation]
{
	\includegraphics[width=4.1cm]{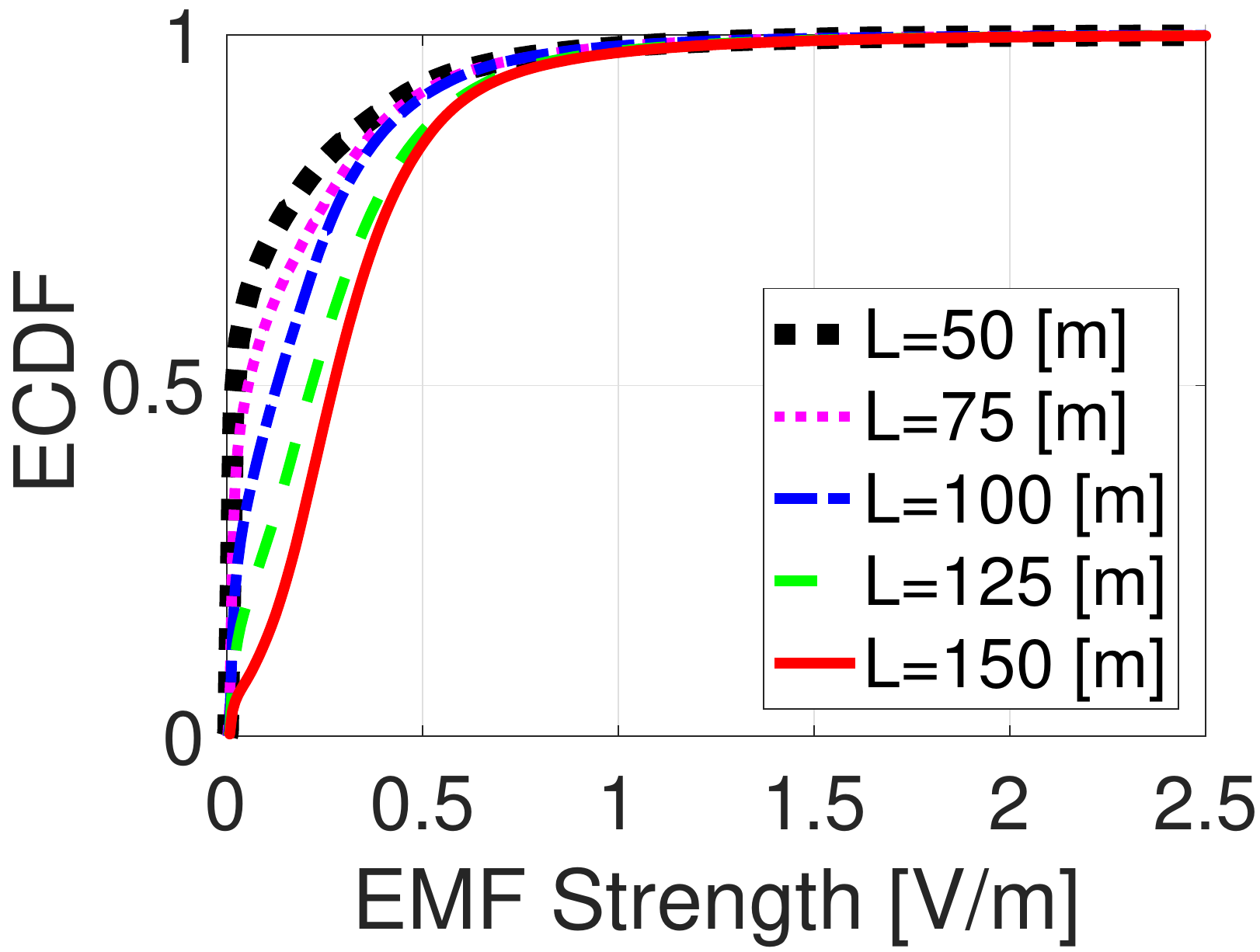}
	\label{fig:emf_eval_var_l}
}
\subfigure[Throughput Evaluation]
{
	\includegraphics[width=4.1cm]{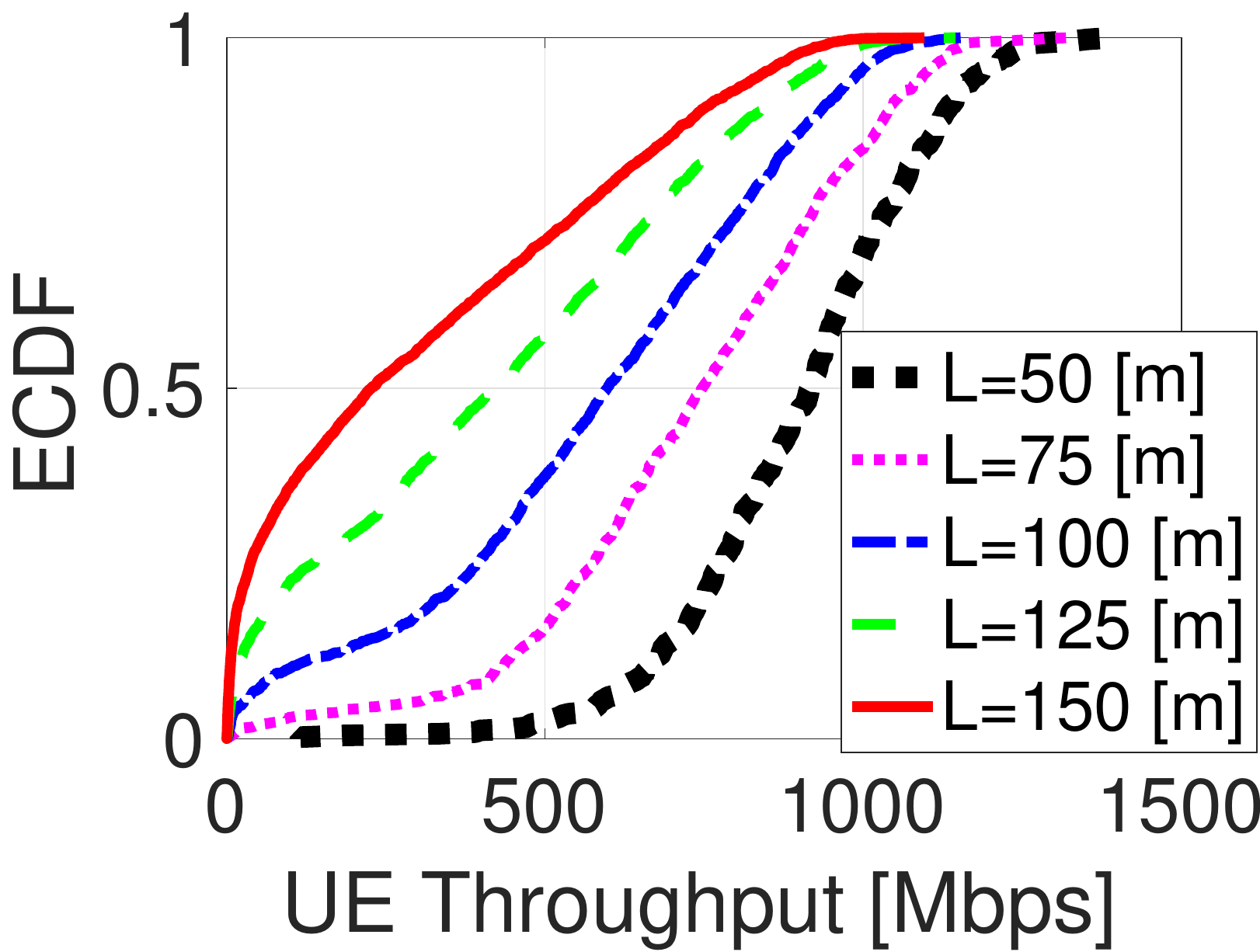}
	\label{fig:throughput_eva_var_l}
}
\caption{EMF and throughput evaluation vs. variation of sector size $L$ (constant \ac{UE} density).}
\label{fig:var_param}
\end{figure}

\textbf{Impact of sector size and \ac{UE} density.} In this part, we move our attention on the impact of pencil beamforming when the size of each sector (parameter $L$) \textit{and} the \ac{UE} density (i.e., number of deployment spots per sector over the sector area) are varied. To this aim, we consider the following cases: \textit{i}) \ac{UE} density increased with $L$, and \textit{ii}) \ac{UE} density kept constant w.r.t. $L$. Focusing on case \textit{i}), we assume to always generate a number of deployment spots per sector equal to $N^{\text{R}}_s=64$, i.e., the number of radiating elements. In this way, the \ac{UE} density is reduced when $L$ is increased. On the other hand, we impose a fixed \ac{UE} density in case \textit{ii}), i.e., the number of deployment spot per sector is proportional to the sector area, with a maximum value (i.e., equal to $N^{\text{R}}_s=64$) achieved for the largest $L$. We then run \textsc{5G-Pencil} for each setting, by assuming the following setting: \textit{i}) only inter-sector interference (i.e., the intra-sector term in Eq.~(\ref{eq:sinr}) is not considered), \textit{ii}) $\epsilon=2$~[m] and \textit{iii}) 20 independent runs for generating the deployment spots.

Fig.~\ref{fig:var_param_dp_constant} reports the \ac{ECDF} of \ac{EMF} and throughput for different values of $L$, by considering the case in which the \ac{UE} density is decreasing with $L$. We remind that, with this setting, the number of deployment spots per sector is always equal to 64. Interestingly, the increase of $L$ triggers a prompt decrease of \ac{EMF} (Fig.~\ref{fig:emf_eval_var_l_constant_dp}), due to the fact that the beam exposure overlapping is reduced (from both the same sector and the neighboring ones). In addition, the throughput tends to be decreased when $L$ is increased (Fig.~\ref{fig:throughput_eva_var_l_constant_dp}), due to the larger distance (and hence worser propagation conditions) that is experienced by the deployment spot w.r.t. the serving sector.

We then move our attention to the investigation of \ac{EMF} and throughput when the \ac{UE} density is kept constant, as shown in Fig.~\ref{fig:var_param}. Interestingly, the \ac{EMF} tends to be increased with $L$ (Fig.\ref{fig:emf_eval_var_l}), in contrast to the previous setting (Fig.~\ref{fig:emf_eval_var_l_constant_dp}). However, we remind that in Fig.~\ref{fig:var_param} we are increasing the number of deployed beams when $L$ is increased, and therefore this setting introduces more radiating sources over the territory. In line with the previous case, the throughput is improved when $L$ is decreased (Fig.~\ref{fig:throughput_eva_var_l}), thanks again to the shorter distance w.r.t. the serving sector. However, by comparing Fig.~\ref{fig:throughput_eva_var_l} and Fig.~\ref{fig:throughput_eva_var_l_constant_dp}, the throughput is better in the former compared to the latter. Such difference is more evident for the lowest values of $L$. In this case, in fact, a reduced number of inter-sector interferers is introduced in Eq.~(\ref{eq:sinr}), thus notably improving the observed throughput values.

\begin{figure}[t]
\centering
\includegraphics[width=8cm]{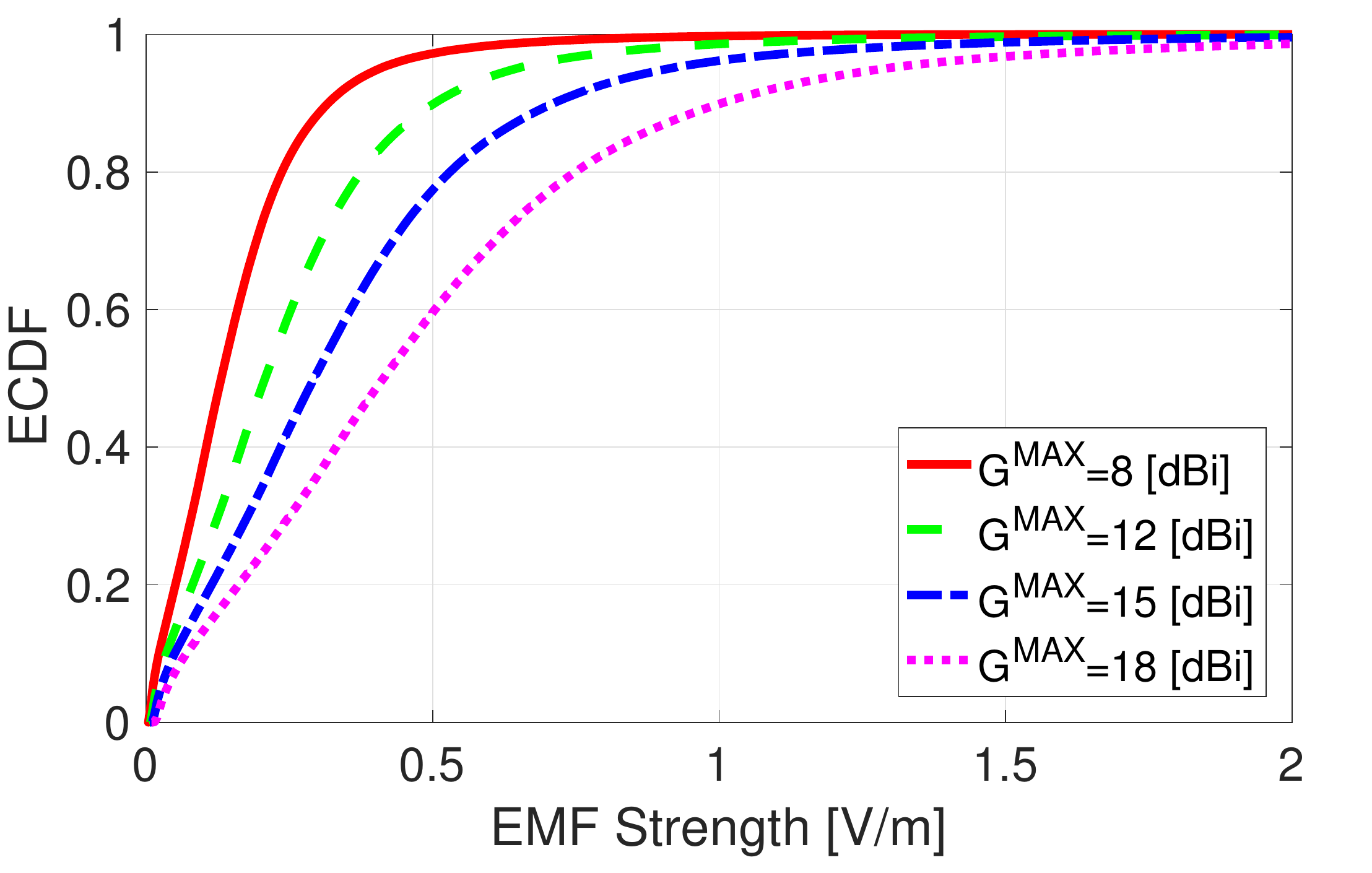}
\caption{\ac{ECDF} of the \ac{EMF} strength vs. the variation of maximum antenna gain $G^{\text{MAX}}$ with pencil beamforming ($\epsilon=2$~[m]).}
\label{fig:Adapt_antenna_gain_var}
\end{figure}

\textbf{Impact of Antenna Gain.} In the final part of our work, we analyze the impact of the maximum antenna gain $G^{\text{MAX}}$ that is used for the \ac{EMF} evaluation. Fig.~\ref{fig:Adapt_antenna_gain_var} reports the \ac{ECDF} of the \ac{EMF} obtained by \textsc{5G-Pencil} for different values of $G^{\text{MAX}}$, by assuming again $\epsilon=2$~[m] and 20 independent runs. As expected, the increase of $G^{\text{MAX}}$ tends to increase the exposure levels. However, the \ac{EMF} increase is overall pretty limited in terms of absolute value, with a maximum \ac{EMF} strength always lower than 2~[V/m] for almost all the measurement spots.

\section{Summary and Future Works}
\label{sec:conclusions}

We have investigated the impact of pencil beamfoming on \ac{EMF} and throughput levels, by designing and evaluating the \textsc{5G-Pencil} framework. Our solution, which operates as an upper layer on top of 5G core functionalities, leverages the localization uncertainty level to tune the direction and width for each traffic beam. We have then coded the presented framework as a publicly-released open-source simulator, in order to compute the \ac{EMF} and the throughput of pencil beamforming in a meaningful set of scenarios. Our results demonstrate that the supposed \ac{EMF} increase associated with pencil beamforming is not supported by scientific evidence. On the contrary, when the tuning of the traffic beams integrates localization information, a strong exposure reduction is observed not only over the deployment spots but also on the whole territory. In addition, the better is the localization uncertainty level, the narrower are the synthesized beams, and consequently the lower is the \ac{EMF} exposure. Eventually, large throughput levels can be achieved when each deployment spot is served by one dedicated traffic beam. 

As future work, we will consider the pencil beam activation/deactivation over space and over time, in order to minimize the exposure (while preserving the user Quality of Service). Such investigation will include e.g., power lock mechanisms to avoid exposure spikes. In addition, the adaptation of the traffic beams to track \ac{UE} mobility over the territory is another avenue of research. Eventually, we will investigate the trade-off between the timing to synthesize a beam and the age of localization information, also in terms of exposure evaluation. Finally, we plan to extend our framework by considering pencil beams over mm-Waves frequencies.

\section*{Acknowledgements}
This work was supported by the European Union's Horizon 2020 Research and Innovation Programme under Grant number 871249. The authors would like to thank Matteo Arciuli for his help in coding the initial version of the simulator.

\bibliographystyle{ieeetr}

\begin{thebibliography}{10}

\bibitem{beammyth}
{\em {5G: A dangerous generation}}.
\newblock Available at
  \url{https://www.downtoearth.org.in/news/science-technology/5g-a-dangerous-generation-63802},
  last accessed on 21th Jan. 2021.

\bibitem{chiaraviglio2020health}
L.~Chiaraviglio, A.~Elzanaty, and M.-S. Alouini, ``Health risks associated with
  5g exposure: A view from the communications engineering perspective,'' {\em
  arXiv preprint arXiv:2006.00944}, 2020.

\bibitem{5gfire}
{\em {77 cell phone towers have been set on fire so far due to a weird
  coronavirus 5G conspiracy theory}}.
\newblock Available at
  \url{https://www.businessinsider.com/77-phone-masts-fire-coronavirus-5g-conspiracy-theory-2020-5?IR=T},
  last accessed on 21th Jan. 2021.

\bibitem{bushberg2020ieee}
J.~Bushberg, C.~Chou, K.~Foster, R.~Kavet, D.~Maxson, R.~Tell, and M.~Ziskin,
  ``{IEEE Committee on Man and Radiation - COMAR Technical Information
  Statement: Health and safety issues concerning exposure of the general public
  to electromagnetic energy from 5G wireless communications networks},'' {\em
  Health Physics}, vol.~119, no.~2, p.~236, 2020.

\bibitem{medicalpress2}
{\em {Does 5G pose health risks? (part 2)}}.
\newblock Available at
  \url{https://www.edn.com/does-5g-pose-health-risks-part-2/}, last accessed on
  22th Oct. 2020.

\bibitem{3gpp.23.273}
3GPP, ``{5G System (5GS) Location Services (LCS); Stage 2},'' Technical
  Specification (TS) 23.273, {3rd Generation Partnership Project (3GPP)}, 7
  2020.
\newblock Version 16.4.0.

\bibitem{roh2014millimeter}
W.~Roh, J.-Y. Seol, J.~Park, B.~Lee, J.~Lee, Y.~Kim, J.~Cho, K.~Cheun, and
  F.~Aryanfar, ``{Millimeter-wave beamforming as an enabling technology for 5G
  cellular communications: Theoretical feasibility and prototype results},''
  {\em IEEE Comm. Magazine}, vol.~52, no.~2, pp.~106--113, 2014.

\bibitem{yu2016load}
B.~Yu, L.~Yang, and H.~Ishii, ``{Load balancing with 3-d beamforming in
  macro-assisted small cell architecture},'' {\em IEEE Transactions on Wireless
  Communications}, vol.~15, no.~8, pp.~5626--5636, 2016.

\bibitem{awada2017simplified}
A.~Awada, A.~Lobinger, A.~Enqvist, A.~Talukdar, and I.~Viering, ``{A simplified
  deterministic channel model for user mobility investigations in 5G
  networks},'' in {\em IEEE ICC, Paris, France}, pp.~1--7, 2017.

\bibitem{ali2019system}
A.~Ali, U.~Karabulut, A.~Awada, I.~Viering, O.~Tirkkonen, A.~N. Barreto, and
  G.~P. Fettweis, ``{System Model for Average Downlink SINR in 5G Multi-Beam
  Networks},'' in {\em IEEE PIMRC, Instabul, Turkey}, pp.~1--6, 2019.

\bibitem{medicalpress}
{\em {'Very low' risk of unknown health hazards from exposure to 5G wireless
  networks}}.
\newblock Available at
  \url{https://medicalxpress.com/news/2020-06-unknown-health-hazards-exposure-5g.html},
  last accessed on 22th Oct. 2020.

\bibitem{NZbeam}
{\em {5G and Health - Ministry of Health, New Zealand}}.
\newblock Available at
  \url{https://www.health.govt.nz/system/files/documents/topic_sheets/5g-and-health-aug19.pdf},
  last accessed on 22th Oct. 2020.

\bibitem{Tho-17}
B.~Thors, A.~Furusk{\"a}r, D.~Colombi, and C.~T{\"o}rnevik, ``{Time-Averaged
  Realistic Maximum Power Levels for the Assessment of Radio Frequency Exposure
  for 5G Radio Base Stations Using Massive MIMO},'' {\em IEEE Access}, vol.~5,
  pp.~19711--19719, 2017.

\bibitem{nasim2019adverse}
I.~Nasim and S.~Kim, ``{Adverse impacts of 5G downlinks on human body},'' in
  {\em IEEE SoutheastCon, Hunstville, Alabama}, pp.~1--6, 2019.

\bibitem{basikolo2019electromagnetic}
T.~Basikolo, T.~Yoshida, and M.~Sakurai, ``{Electromagnetic Field Exposure
  Evaluation for 5G in Millimeter Wave Frequency Band},'' in {\em IEEE
  International Symposium on Antennas and Propagation and USNC-URSI Radio
  Science Meeting, Atlanta, Georgia}, pp.~1523--1524, 2019.

\bibitem{loh2020assessment}
T.~H. Loh, F.~Heliot, D.~Cheadle, and T.~Fielder, ``{An Assessment of the Radio
  Frequency Electromagnetic Field Exposure from A Massive MIMO 5G Testbed},''
  in {\em European Conference on Antennas and Propagation (EuCAP), Copenhagen,
  Denmark}, pp.~1--5, IEEE, 2020.

\bibitem{xu2020emf}
B.~Xu, D.~Colombi, and C.~T{\"o}rnevik, ``{EMF exposure assessment of massive
  MIMO radio base stations based on traffic beam pattern envelopes},'' in {\em
  European Conference on Antennas and Propagation (EuCAP), Copenhagen,
  Denmark}, pp.~1--5, IEEE, 2020.

\bibitem{adda2020theoretical}
S.~Adda, T.~Aureli, S.~D’elia, D.~Franci, E.~Grillo, M.~D. Migliore,
  S.~Pavoncello, F.~Schettino, and R.~Suman, ``{A Theoretical and Experimental
  Investigation on the Measurement of the Electromagnetic Field Level Radiated
  by 5G Base Stations},'' {\em IEEE Access}, vol.~8, pp.~101448--101463, 2020.

\bibitem{3GPP:TS:22.071:V16.0.0}
3GPP, ``Location services {(LCS)},'' Technical Specification (TS) 22.071, {3rd
  Generation Partnership Project (3GPP)}, July 2020.
\newblock Version 16.0.0.

\bibitem{3GPP:TS:23.032:V16.0.0}
3GPP, ``Universal geographical area description {(GAD)},'' Technical
  Specification (TS) 23.032, {3rd Generation Partnership Project (3GPP)}, July
  2020.
\newblock Version 16.0.0.

\bibitem{iturec70}
{\em {ITU-T K70 Mitigation techniques to limit human exposure to EMFs in the
  vicinity of radiocommunication station}}.
\newblock Available at \url{https://www.itu.int/rec/T-REC-K.70/en}, last
  accessed on 26th Oct. 2020.

\bibitem{iturec91}
{\em {ITU-T K.91: Guidance for assessment, evaluation and monitoring of human
  exposure to radio frequency electromagnetic fields}}.
\newblock Available at \url{https://www.itu.int/rec/T-REC-K.91-201911-I/en},
  last accessed on 26th Oct. 2020.

\bibitem{Tornevik}
{\em {Impact of EMF limits on 5G network roll-out}}.
\newblock Available at
  \url{https://www.itu.int/en/ITU-T/Workshops-and-Seminars/20171205/Documents/S3_Christer_Tornevik.pdf},
  last accessed on 27th Oct. 2020.

\bibitem{international2020guidelines}
{{International Commission on Non-Ionizing Radiation Protection}}, ``Guidelines
  for limiting exposure to electromagnetic fields (100 khz to 300 ghz),'' {\em
  Health Physics}, vol.~118, no.~5, pp.~483--524, 2020.

\bibitem{5gpencil}
{\em {5GPencil Simulator 1.1}}.
\newblock Available at \url{https://tinyurl.com/5GPencil}, last accessed on
  21th Jan. 2021.

\bibitem{gnbdatastheets}
{\em {Hamilton 3.5 GHz 8x8 MIMO Panel Antenna}}.
\newblock Available at
  \url{https://halberdbastion.com/products/antenna-catalogue/hamilton-35-ghz-8x8-mimo-panel-antenna},
  last accessed on 27th Oct. 2020.

\bibitem{italianauction}
{\em {Italian 5G spectrum auction}}.
\newblock Available at
  \url{https://5gobservatory.eu/italian-5g-spectrum-auction-2/}, last accessed
  on 15th Jan. 2021.

\bibitem{pathlossmodel}
{\em {3GPP TR 38.901 version 16.1.0 Release 16}}.
\newblock Available at
  \url{https://www.etsi.org/deliver/etsi_tr/138900_138999/138901/16.01.00_60/tr_138901v160100p.pdf},
  last accessed on 15th Jan. 2021.

\bibitem{noisesetting}
{\em {ITU-R SG05 Contribution 57: Guidelines for evaluation of radio interface
  technologies for IMT-2020}}.
\newblock Available at
  \url{https://www.health.govt.nz/system/files/documents/topic_sheets/5g-and-health-aug19.pdf},
  last accessed on 15th Jan. 2021.

\bibitem{3GPP:TS:22.261:V18.1.0}
3GPP, ``Service requirements for the {5G} system,'' Technical Specification
  (TS) 22.261, {3rd Generation Partnership Project (3GPP)}, Dec. 2020.
\newblock Version 18.1.0.

\bibitem{beamformers}
{\em {Beamformers Explained}}.
\newblock Available at
  \url{https://www.commscope.com/globalassets/digizuite/542044-Beamformer-Explained-WP-114491-EN.pdf},
  last accessed on 20th Jan. 2021.

\bibitem{fazliu2020mmwave}
Z.~L. Fazliu, F.~Malandrino, C.~F. Chiasserini, and A.~Nordio, ``Mmwave beam
  management in urban vehicular networks,'' {\em IEEE Systems Journal}, 2020.

\end{thebibliography}

\end{document}